\documentclass[aps,12pt, prb, preprint]{revtex4}
\usepackage{srctex}
\usepackage{amsmath}
\usepackage{amsfonts}
\usepackage{amssymb}
\usepackage{graphicx}
\usepackage[dvips,dvipsnames]{xcolor}

\newcommand{\kBoltzmann}{k_{B}}

\begin{document}

\title{Revisiting the concept of chemical potential in classical and quantum gases: A perspective from Equilibrium Statistical Mechanics}

\author{F.J. Sevilla}
\email{fjsevilla@fisica.unam.mx}
\affiliation{Instituto de F\'{\i}sica, UNAM, Apdo. Postal 20-364,
01000 M\'exico D.F., MEXICO}

\author{L. Olivares-Quiroz}
\email{luis.olivares@uacm.edu.mx}
\thanks{corresponding author}
\affiliation{Universidad Autonoma de la Ciudad de Mexico. \\ Av La Corona 320 Loma Alta. Gustavo A Madero CP 07160. Mexico D.F. MEXICO}

\begin{abstract}

In this work we revisit the concept of chemical potential $\mu$ in both classical and quantum gases from a perspective of Equilibrium Statistical Mechanics (ESM). Two new results regarding the equation of state $\mu=\mu(n,T)$, where $n$ is the particle density and $T$ the absolute temperature, are given for the classical interacting gas and for the weakly-interacting quantum Bose gas. In order to make this review self-contained and adequate for a general reader we provide all the basic elements in an advanced-undergraduate or graduate statistical mechanics course required to follow all the calculations. We start by presenting a calculation of $\mu(n,T)$ for the classical ideal gas in the canonical ensemble. After this, we consider the interactions between particles and compute the effects of them on $\mu(n,T)$ for the van der Waals gas. For quantum gases we present an alternative approach to calculate the Bose-Einstein (BE) and Fermi-Dirac (FD) statistics. We show that this scheme can be straightforwardly generalized to determine what we have called Intermediate Quantum Statistics (IQS) which deal with ideal quantum systems where a single-particle energy can be occupied by at most $j$ particles with $0 \leqslant  j \leqslant  N$ with $N$ the total number of particles. In the final part we address general considerations that underlie the theory of weakly interacting quantum gases. In the case of the weakly interacting Bose gas, we focus our attention to the equation of state $\mu=\mu(n,T)$ in the Hartree-Fock mean-field approximation (HF) and the implications of such results in the elucidation of the order of the phase transitions involved in the BEC phase for non-ideal Bose gases.

\end{abstract}

\maketitle

\section{Introduction}
Chemical potential has proven to be a subtle concept in thermodynamics and statistical mechanics since its appearance in the classical works of J.W. Gibbs \cite{GibbsW}. Unlike thermodynamic concepts such as temperature $T$, internal energy $E$ or even entropy $S$, chemical potential $\mu$ has acquired, justified or not, a reputation as a concept not easy to grasp even for the experienced physicist. Gibbs introduced  chemical potential within the context of an extense and detailed exposition on the foundations of what is now called statistical mechanics. In his exposition he considers how to construct an ensemble of systems which can exchange particles with the surroundings. In such description, $\mu$ appears as a constant required to provide a necessary closure to the corresponding set of equations \cite{GibbsW}. A fundamental connection with thermodynamics is thus achieved by observing that the until-then unknown constant $\mu$ is indeed related, through first derivatives, to standard thermodynamic functions like the Helmholtz free energy $F=E-TS$ or the Gibbs thermodynamic potential $G=F+pV$.  In fact, $\mu$ appeared as a conjugate thermodynamic variable to the number $N$ of particles in the same sense as pressure $p$ is a conjugate variable to volume $V$. The procedure outlined above and described with detailed elegance by J.W. Gibbs defines the essence of the chemical potential in statistical mechanics and thermodynamics.\\

The link provided by Gibbs to define chemical potential in terms of thermodynamic variables is certainly a master piece, however,  a direct physical interpretation might still be elusive. Consider for example, two of the most used definitions of $\mu$ in equilibrium thermodynamics \cite{Callen}, \cite{LGCS}, i.e.,
\begin{equation}\label{Equation22Feba}
\mu= \left( \frac{\partial F}{\partial N} \right)_{T,V}= \left(\frac{\partial E}{\partial N}\right)_{S,V},
\end{equation}
where $V$ is the system volume. As can be readily seen, the first definition in terms of $F$'s derivative implies that we can obtain $\mu$ as a measure of the change of $F$ with respect to the number $N$ of particles \emph{at constant volume and temperature}. It is straightforward to imagine a closed box of volume $V$ where we can add or subtract particles and observe changes in the free energy of the system. However, depicting such situation when both volume $V$ and temperature $T$ are kept fixed may require a higher degree of physical intuition recalling that any particle added to the system will provide some additional amount of energy either in the form of potential or kinetic energy. Let us consider the second term in Eq (\ref{Equation22Feba}). This thermodynamic definition suggests that $\mu$ can be measured as the change of internal energy $E$ with respect to the number $N$ of particles but this time keeping constant entropy $S$ and volume $V$. What exactly must be understood by adding particles keeping entropy constant? Recall that each particle added to the system brings an increase in the number of configurations available to the overall system and therefore an increase of entropy would be our first intuitive expectation. In \emph{equilibrium statistical mechanics} (ESM) the procedure to present $\mu$ is very much based on the approach suggested by Gibbs in his classic works. The central idea behind ESM is a many-variable minimization process in order to obtain a distribution function $\{n_q \}$  corresponding to an extremal, minimum or maximum, of the thermodynamic variables $F$ or $S$ respectively. In this context $\mu$ appears as a mathematical auxiliary quantity identified with a Lagrange multiplier that minimizes/maximizes a physical quantity.\\

In this work we present a discussion on the concept of chemical potential from a perspective of ESM and how it emerges from physical considerations in both classical and quantum gases. Our main focus is to present to undergraduate and graduate students a self-contained review on the basic elements that give rise to an understanding of $\mu.$ In order to achieve this goal we shall proceed as follows. Section II carries out a detailed calculation of $\mu$ in the case of an ideal classical gas. Using a method proposed by van Kampen \cite{vanKampenPhysica61} we include the effect of interactions and calculate $\mu$ for the van der Waals gas. Section III deals with ideal quantum gases. We introduce calculations by giving a general discussion on the temperature scale at which quantum effects are expected to contribute significantly. A general description, based on a simple, but novel method to compute the average number of particles $\langle n_{k}\rangle$ that occupy the single-particle energy level $\epsilon_{k}$ for boson and fermions is introduced. We also provide a formal calculation for $\langle n_k \rangle_j$ when the energy levels can be occupied at most by $j$ particles, where $0 \leqslant  j \leqslant  N$. We call the resulting statistics the Intermediate Quantum Statistics (IQS) of order $j,$ which generalizes the BE and FD statistics which are obtained for $j\to\infty$ and $j=1$, respectively.
Finally in Section IV we go a step further to consider the behavior of $\mu$ as a function of particle density $n$ and temperature $T$ for the weakly interacting quantum Bose and Fermi gas. The former system has been under intense research lately since it is the standard theoretical model to describe Bose-Einstein Condensation (BEC) in ultracold alkali atoms \cite{Pethick}, and as we outlined here, the knowledge of $\mu=\mu(n,T)$ is of fundamental importance since it contains valuable information on the nature of the phase transition involved.\\

Our intention is not to provide an exhaustive treatment of the chemical potential in ESM, instead, our contribution intends to integrate previous well known results within a physically intuitive framework, and at the same time to provide some new results that might be interesting to the reader that complement and enhance a broader view of the subject. We kindly invite to the interested reader to study several excellent textbooks \cite{pathria, ReifBook, HuangBook, LandauLifshitz} and reviews \cite{cook_ajp95}, \cite{Tobochnik}, \cite{baierlein, job2006, KaplanT}  that have been written on the subject in the recent past.

\section{\label{sectII}Chemical potential I: the classical gas}
In order to discuss chemical potential for the ideal classical gas we shall address some basic considerations  and implications that $\mu$ must satisfies according to general thermodynamic principles. Although we present them at this point, its validity goes beyond the classical ideal gas. Fundamental postulate in equilibrium thermodynamics \cite{Callen} assures that for a given system there is a function called the entropy $S$ defined only for equilibrium states which depends on volume $V$, internal energy $E$ and number of particles $N$, i.e, $S=S(E,V,N)$. Thus, an infinitesimal change $d S$ between two equilibrium states can be written as
\begin{equation}\label{Equation2}
dS= \left( \frac{\partial S}{\partial E}\right)_{V,N} dE +\left( \frac{\partial S}{\partial V}\right)_{E,N} dV+\left( \frac{\partial S}{\partial N}\right)_{V,E} dN.
\end{equation}
Using the \emph{first law of thermodynamics} $dE=TdS-pdV$ we can relate the partial derivatives that appear in Eq (\ref{Equation2}) with standard thermodynamic variables temperature $T$ and pressure $p$. A simple inspection points out that
\begin{eqnarray}\label{Equation3}
\left( \frac{\partial S}{\partial E}\right)_{V,N} &=& \frac{1}{T} \nonumber \\
\left( \frac{\partial S}{\partial V}\right)_{E,N} &=& \frac{p}{T}. \\
\end{eqnarray}
Such identification suggests that we must add to the First Law a suitable thermodynamic variable that will play the role of a conjugate variable to the number of particles $N$ and that will allow the connection to $\left( \partial S / \partial N \right)_{V,E}$, just like $T$ is conjugated to the entropy $S$ and $p$ is to the volume $V$. Thus, if we allow the exchange of particles, we can write the First Law as $ dE=Tds-pdV+ \mu dN$ and hence
\begin{equation}\label{Equation4}
- T \left( \frac{\partial S}{\partial N}\right)_{V,E} = \mu.
\end{equation}
Eq. (\ref{Equation4}) provides additional information on the nature of chemical potential complementing Eqs. (\ref{Equation22Feba}). This tells us that $\mu$ is a negative quantity if  entropy increases with the number of particles by keeping energy $E$ and volume $V$ constant. Though it is intuitive that $S$ increases as $N$ increases, it is not the case under the restrictions of $E$ and $V$ constant.  On the other hand, Eq. (\ref{Equation4}) also admits the possibility that $\mu > 0$, however, as we show below for the ideal Fermi gas and the weakly interacting Bose gas in sections III- IV, respectively, this is true only as a result of quantum effects.

\subsection{The classical ideal gas}

To determine $\mu$ as a function of $(E,V,N)$ we shall make use of the fundamental equation $S=S(E,V,N)$ and Eq. (\ref{Equation4}). ESM ensures that the macroscopic variable entropy $S$ is related to a microscopic quantity $\Omega(E,V,N)$ which represents the number of microstates available to the system consistent with the macroscopic restrictions of constant $E$, $V$ and $N$. Such connection is given by  $S=k_{B}\ln\Omega(E,V,N)$  where $k_B$ is the Bolztmann's constant. With this considerations, $\Omega$ is given by
\begin{multline}\label{Omega}
\Omega(E,V,N)=\frac{1}{N!h^{3N}}\int\cdots\int \delta(E-H)d^{3}{\bf r}_{1}\, d^{3}{\bf p}_{1}^{3} d^{3}{\bf r}_{3N}\, d^{3}{\bf p}_{3N}
\end{multline}
where $H=\sum_{i=1}^{3N}p_{i}^{2}/2m$ is the Hamiltonian for a system of $N$ free particles and $1/N!$ corresponds to the Gibb's correction factor. Since $H$ is ${\bf r}$-independent and spherically symmetric respect to momentum coordinates ${\bf p}_{i}$ we can write Eq. (\ref{Omega}) as
\begin{multline}
\Omega(E,V,N)=\frac{V^{N}}{N!h^{3N}}\frac{2\pi^{3N/2}}{\Gamma(3N/2)}\times
\int_{0}^{\infty}dP\, P^{3N-1} \delta(E-P^{2}/2m)
\end{multline}
where the change of variable $P\equiv\sum_{i=1}^{3N}p_{i}^{2}$ has been made and the hyper-volume element in 3$N$ dimensions with coordinates $p_{i}$  given by $d\Pi=2\pi^{3N/2}P^{3N-1}dP,$ with $P$ defined as before, has been used. Then,
\begin{equation}\label{Equation10}
\Omega(E,V,N)=\frac{1}{N!}\frac{V^{N}}{h^{3N}}\left(2\pi m\right)^{3N/2}\frac{E^{3N/2-1}}{(3N/2-1)!}.
\end{equation}
In the limit where $N\gg1,$ Eq. (\ref{Equation10}) can be written as
\begin{equation}\label{Equation11}
\Omega(E,V,N)=\frac{1}{N!}\frac{V^{N}}{h^{3N}}\left(2\pi m\right)^{3N/2}\frac{E^{3N/2}}{(3N/2)!},
\end{equation}
Thus, entropy $S$ can be readily calculated. After using  Stirling's approximation one obtains
\begin{equation}\label{Equation13}
S=k_{B}N\left\{\ln\frac{V}{N}+\frac{3}{2}\ln\left[\frac{mE}{3\pi\hbar^{2}N}\right]+\frac{5}{2}\right\}.
\end{equation}
Substitution of  (\ref{Equation13}) into (\ref{Equation4}) leads to the well known result for the chemical potential for the ideal classical gas
\begin{equation}\label{Equation14}
\mu_{ideal}=-k_{B}T\, \ln\left[\frac{V}{N}\left(\frac{mk_{B}T}{2\pi\hbar^{2}}\right)^{3/2}\right],
\end{equation}
where the relation $E=\frac{3}{2}Nk_{B}T$ has been used.\\

Eq. (\ref{Equation14}) has an interesting interpretation in terms of the average distance between particles  $l\equiv (V/N)^{1/3}$  and the thermal-wavelength $\lambda_{T}= h / \sqrt{2 \pi m k_B T}$ (see section \ref{sectIII} for a larger discussion). As it can be seen from Eq. (\ref{Equation14}), $\mu = -k_B T \ln \left[ {l^{3}/ \lambda_{T}^{3}} \right]$ from which a physical interpretation can be easily harnessed. The sign of the chemical potential is determined then by the ratio $l / \lambda_{T}$. In the high-temperature limit, when the quantum effects are small and the wave nature of particles is negligible in comparison to $l$, i.e, $\lambda_{T} \ll l$, $\mu$ is negative and the system can be regarded as formed of idealized punctual particles that can be distinguished, in principle, one from each other. This picture corresponds to the ideal classical gas. This interpretation opens up the possibility that in the quantum regime, $\lambda_{T}\sim\l,$ $\mu$ could acquire positive values.\\

We can gain additional information if we consider the discrete version of Eq (\ref{Equation4}), namely
\begin{equation}
\mu=-T(\Delta S)_{E,V}
\end{equation}
where
\begin{equation}\label{Equation12}
(\Delta S)_{E,V}=k_{B}\ln\frac{\Omega(E,V,N+1)}{\Omega(E,V,N)}.
\end{equation}
Notice that Eq(\ref{Equation12}) gives the sign of the chemical potential when one particle is exactly added to the system keeping $E$ and $V$ constant. Substitution of Eq  (\ref{Equation10}) into Eq (\ref{Equation12}) yields
\begin{equation}\label{EquationFeb28A}
\frac{\Omega(E,V,N+1)}{\Omega(E,V,N)}=\frac{V}{(N+1)}\frac{(\frac{3}{2}N-1)!}{(\frac{3}{2}N+\frac{1}{2})!}\left(\frac{m}{2\pi\hbar^{2}}\right)^{3/2}E^{3/2}.
\end{equation}
For $N \gg 1 $,
\begin{equation}\label{OmegaRatio}
\frac{\Omega(E,V,N+1)}{\Omega(E,V,N)}\simeq\frac{2}{3}\frac{V}{N}e^{1/2}\left(\frac{mE}{3\pi\hbar^{2}N}\right)^{3/2},
\end{equation}
where  Stirling's approximation has been used. In this representation, $\mu$ goes essentially as the logarithm of the ratio between the energy per particle $E/N$ and the energy  $ \varepsilon = \hbar^{2}/2m(V/N)^{2/3}$ of a quantum particle confined in a box of volume $V$. The condition $E/N \gg \varepsilon $ guarantees the classical character of the system assigning a negative value to the chemical potential.

\subsection{The effects of interactions}

It has been shown in  previous section that  chemical potential for the ideal classical gas is a negative quantity for the whole temperature region where quantum effects can be neglected. In order to enhance our intuition on the nature of chemical potential we shall address the calculation of $\mu$ in the case of a classical gas with pairwise interactions between particles. For a system of $N$ particles, the total partition function  $Z_{N}$ can be written as \cite{ReifBook,HuangBook,pathria}
\begin{equation}\label{RealGasZ}
Z_{N}=\frac{1}{N!}\left(\frac{mk_{B}T}{2\pi\hbar^{2}}\right)^{3N/2} Q_{N},
\end{equation}
where
\begin{equation}\label{Qconfigurational}
Q_{N}=\int e^{-\beta(v_{1,2}+v_{1,3}+\ldots+v_{N-1,N})}d^{3}{\bf r}_{1}\ldots d^{3}{\bf r}_{N}
\end{equation}
is known as the configurational integral. In Eq. (\ref{Qconfigurational}) $v_{i,j}\equiv v(\vert{\bf r}_{i}-{\bf r}_{j}\vert)$ is the interaction energy between the $i$-th and $j$-th particles and $\beta=(k_{B}T)^{-1}$ as usual.\\

A simple method to evaluate $Q_{N}$ has been given by van Kampen in Ref. \cite{vanKampenPhysica61}. In such work, it is suggested that the average of $e^{-\beta v_{1,2}}e^{-\beta v_{1,3}}\cdots e^{-\beta v_{N-1,N}}$ over all possible configurations of  particle's positions can be identified exactly as the ratio $Q_{N}/V^{N}$. Then, the configurational partition function $Q_N$ can be expressed as \cite{vanKampenPhysica61}
\begin{equation}
Q_{N}=V^{N}\exp\left\{N\sum_{k=1}^{\infty}\left(\frac{N}{V}\right)^{k}\frac{B_{k}}{k+1}\right\},
\label{RealGasZ2}
\end{equation}
where the coefficients $B_k$ are given by
\begin{equation}\label{B_k}
B_{k}\equiv \frac{V^{k}}{k!}\sum_{\{k\}}\int\cdots\int\prod_{i<j}\left(e^{-\beta v_{i,j}}-1\right)d{{\bf r}_{1}}\cdots d{{\bf r}_{k}},
\end{equation}
and the sum is taken over all {\it irreducible} terms that involve $k$-particle position coordinates (see  Appendix for more details). The total partition function $Z_{N}$ is then given by
\begin{equation}
Z_{N}=\frac{V^{N}}{N!}\left(\frac{mk_{B}T}{2\pi\hbar^{2}}\right)^{3N/2}\exp\left\{N\sum_{k=1}^{\infty}\left(\frac{N}{V}\right)^{k}\frac{B_{k}}{k+1}\right\},
\end{equation}\\
and from this, the Helmholtz free energy  $F$ by
\begin{multline}
F=-Nk_{B}T\ln\left[\frac{V}{N}\left(\frac{mk_{B}T}{2\pi\hbar^{2}}\right)^{3/2}\right]-Nk_{B}T\left[1+\sum_{k=1}^{\infty}\left(\frac{N}{V}\right)^{k}\frac{B_{k}}{k+1}\right].
\end{multline}
The chemical potential $\mu$ can be obtained readily as
\begin{equation}\label{mu_realgas}
\mu=\mu_{ideal}-\kBoltzmann T \sum_{k=1}^{\infty}\left(\frac{N}{V}\right)^{k}B_{k}.
\end{equation}
Eq. (\ref{mu_realgas}) gives $\mu$ for the classical interacting gas as a series of powers in the particle density $(N/V )^{k}$. The physical implications are clear, interactions shift the value of the chemical potential from the ideal case. If there is no interactions at all, then $B_{k}=0$ for all $k$ and $\mu= \mu_{ideal}$. In spite of the generality of expression (\ref{mu_realgas}), in practice, calculation of $B_{k}$ for $k > 2$ is rather cumbersome. However, for enough dilute systems, i.e, $N/V\ll 1$, we may consider only the first term of Eq. (\ref{mu_realgas}) as a valid approximation. Thus, at first order in $N/V$ we have $\mu=\mu_{ideal}-\kBoltzmann T(N/V)B_{1},$ where $B_{1}$ depends on the specific interatomic potential between particles. In order to obtain quantitative results about the effects of interactions on the chemical potential we consider the van der Waals gas as a specific example.

\begin{figure}[t]
\centering
\includegraphics[width=0.75\textwidth]{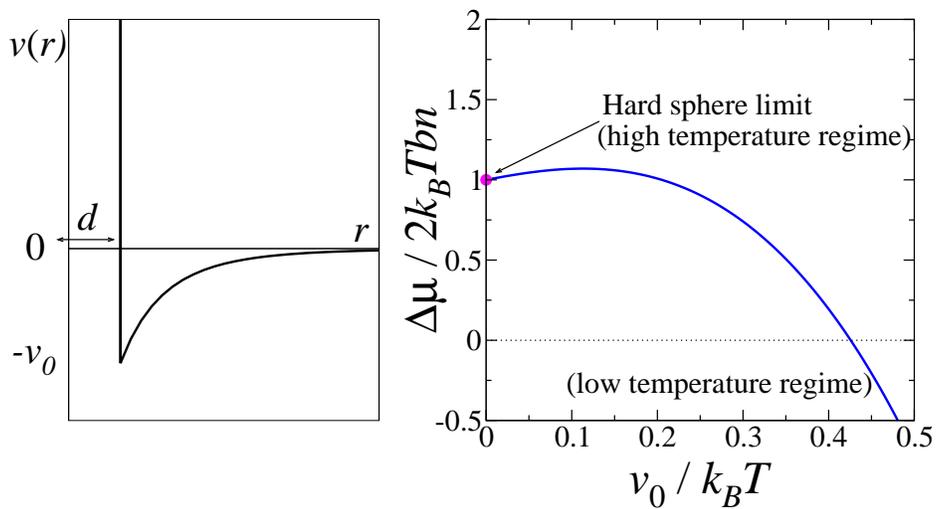}
\caption{Left panel shows the model potential given by expression (\ref{vanWpotential}) as an approximation to the more realistic Lennard-Jones potential. In the right panel we present the change in chemical potential (\ref{Deltamu}) as function of the ratio of the energy that characterizes the interacting potential $v_{0}$ to the thermal energy.}
\label{Fig.DeltaMu}
\end{figure}

Let us consider, for simplicity, the commonly-used pairwise interaction potential
\begin{equation}\label{vanWpotential}
v(r)=\left\{\begin{array}{lr}
\infty & \hbox{for}\,\,\, r<d\\
-v_{0}(d/r)^{6}& \hbox{for}\,\,\, r\ge d,
\end{array}
 \right.
\end{equation}
that approximates the semi-empirical Lennard-Jones potential $v(r)=v_{0}\left[(d/r)^{12}-2(d/r)^{6}\right],$ $v_{0}$ is the minimum interaction energy between a pair of particles and $d$ their separation at which such energy takes place. For this interaction model, $B_{1}$ can be evaluated exactly as follows. By taking advantage of the spherical symmetry of the problem we can write  $B_{1}=\int (e^{-\beta v({\bf r})}-1)d{\bf r}=4\pi\int_{0}^{\infty}r^{2}(e^{-\beta v(r)}-1)dr$. Then, by splitting the last integral into one integral from $0$ to $d$ plus a second one from $d$ to $\infty$ and using the fact that $v(r)\rightarrow\infty$ for $0< r< d$, we get
\begin{equation}\label{Equation1Marzoa}
B_{1}=4\pi \left[\int_{d}^{\infty}(e^{\beta v_{0}(d/r)^{6}}-1)r^{2}dr-\frac{d^{3}}{3}\right].
\end{equation}
The integral in Eq. (\ref{Equation1Marzoa}) can be evaluated directly by using the Taylor series of the exponential function. After integrating term by term we have
\begin{equation}\label{EquationMarzo1b}
B_{1}=\frac{4}{3}\pi d^{3}\left[\sum_{n=1}^{\infty}\frac{(\beta v_{0})^{n}}{(2n-1)n!}-1\right].
\end{equation}
It is possible to go a step further in order to write Eq. (\ref{EquationMarzo1b}) in terms of elementary functions, certainly, the infinite sum can be written as $\sum_{n=1}^{\infty}x^{n}/(2n-1)n!=1-e^{x}+(\pi x)^{1/2}\hbox{Erfi}(x^{1/2})$, where Erfi$(z)=-i$ Erf$(iz)$ denotes the imaginary error function. A simple expression for the correction of $\mu_{ideal}$ due to interactions, defined as $\Delta\mu\equiv\mu-\mu_{ideal},$ can be obtained for temperatures such that $k_{B}T\gg v_{0}$, since only the first term in the series expansion in expression (\ref{EquationMarzo1b}) is needed, with this approximations and recalling that $l=(V/N)^{1/3}$ we have
\begin{equation} \label{Deltamu}
\Delta\mu \simeq k_{B}T\frac{4}{3}\pi \left(\frac{d}{l}\right)^{3}\left(1-\frac{v_{0}}{k_{B}T}\right)>0
\end{equation}
in agreement with Monte Carlo calculations obtained previously by other authors \cite{Adams1974}.
For even higher temperatures, $v_{0}/k_{B}T\approx0,$  and then it is the hardcore repulsion of the inter-particle interaction (\ref{vanWpotential}) what governs the dynamics of the gas. In this limit the system corresponds to a hard-sphere gas thus giving $\Delta\mu=k_{B}T\frac{4}{3}\pi (d/l)^{3}$ \cite{Adams1974}. For temperatures smaller than $v_{0}/k_{B},$ $\Delta\mu$ becomes negative (see Fig. \ref{Fig.DeltaMu}), but this should not be considered correct since at such temperatures we are out of the classical regime and quantum corrections must be taken into account. In terms of the parameters $a$ and $b$ of the standard van der Waals equation of state \begin{equation}\label{vdwaalsequation}
\left(p+\frac{N^{2}}{V^{2}}a\right)(V-Nb)=Nk_{B}T,
\end{equation}
the chemical potential for the van der Waals gas can be written as
\begin{equation}
\mu=\mu_{ideal}-2\frac{N}{V}\left(a-k_{B}T b\right)
\end{equation}
where  $a=v_{0}b=v_{0}\frac{2}{3}\pi d^{3}$. Table I presents some standard values for the $a$ and $b$ values for different gases \cite{DatosVanderWaals}. The interested reader may find useful to see how the calculation just presented works, by using other interaction potentials $v_{i,j}$ between particles.

A relation of $\Delta\mu$ with the work $W({\bf r})$ required to bring an additional particle to the system from infinity to position ${\bf r},$ has been shown by Widom \cite{Widom} as
\begin{equation}\label{WidomEquivalence}
\exp\left(-\Delta\mu/k_{B}T \right)=\langle\exp(-W ({\bf r}) /k_{B}T)\rangle,
\end{equation}
where $\langle ..\rangle$ denotes the canonical-ensemble average. On the other hand, it seems intuitive to expect $W({\bf r})$ to be larger for a gas with repulsive interactions than for the ideal gas, thus, by using Widom's equivalence Eq (\ref{WidomEquivalence}) we may conclude that repulsive interactions yields $\Delta \mu >0.$
\begin{table}\label{Tablevdw}
\centering
\caption{Values of the van der Waals parameters $a$ and $b$ for some substances are given. With these values the ratio $v_{0}/k_{B}T_{R}$ is computed, where $T_{R}$ denotes the room temperature.}
\begin{tabular}{|l|c|c|c|}
\hline
Substance & $a$ & $b$ &$v_{0}/k_{B}T_{R}$\\
\hline
Helium & 0.0346 & 0.0238 & 0.0603 \\
Neon & 0.208 & 0.0167 & \\
Hydrogen & 0.2452 & 0.0265 & 0.384 \\
Oxygen & 1.382 & 0.0319 & 1.796 \\
Water & 5.537 & 0.0305 & 7.527 \\
\hline
\end{tabular}
\end{table}

\section{\label{sectIII}Chemical Potential II: Quantum Ideal Gases}

\begin{figure}
\includegraphics[width=0.5\textwidth]{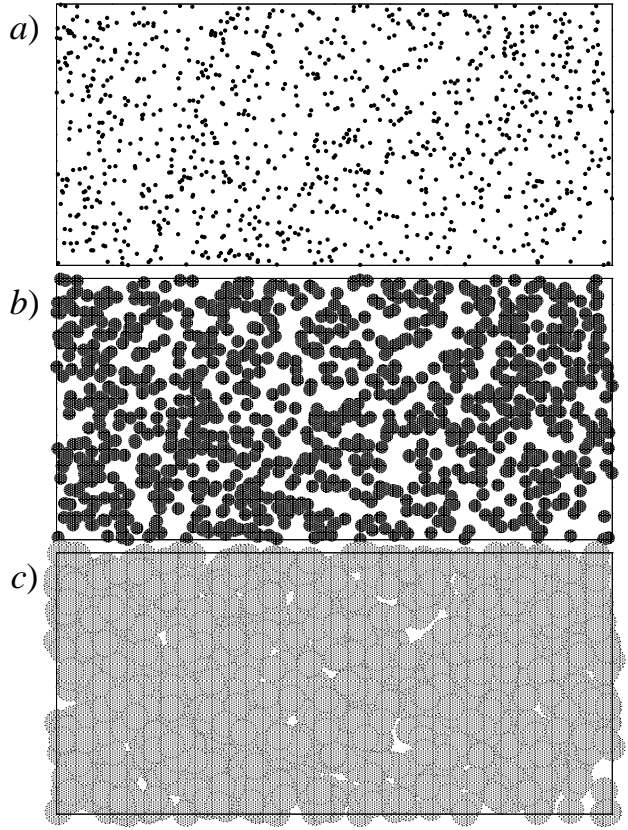}
\caption{In the high temperature limit \emph{a}), quantum statistical correlations measured by the thermal wavelength $\lambda$ are much smaller than the average separation between atoms $l$. \emph{b}) As temperature is lowered, quantum effects start becoming important when relation (\ref{DegeneracyCondition}) holds. At even lower temperatures \emph{c}) indistinguishability is dominant.}
\label{FromClass2Quantum}
\end{figure}

At low temperatures gases depart from their classical behavior since quantum properties of matter emerge. One of the main properties washed away in the quantum regime is distinguishability. In the classical picture, we can in principle label and tag any of the particles, but no longer in the quantum regime \cite{ALeggettBook}. This property has profound consequences in the number of different microstates available to the system. In general, classical systems will have more microstates since permutations among particles result in different configurations due to distinguishability. Quantum systems on the other hand display a smaller number of different configurations. In addition to indistinguishability, quantum gases exhibit another remarkable property. L. de Broglie suggested that any material particle with mass $m$ and velocity $v$ should have a corresponding wavelength $\lambda$ given by
\begin{equation}\label{EquationSectionIIa}
\lambda=\frac{h}{p}=\frac{2 \pi \hbar}{mv},
\end{equation}
where $h$ is the Planck's constant and $p$ the momentum of the particle. Given the fact that a particle with kinetic energy $m v^2/2$ has an associated temperature $T$, it is possible to write down an expression for a thermal de Broglie wavelength $\lambda_{T}$ as
\begin{equation}\label{EquationSectionIIb}
\lambda_T=\frac{h}{\sqrt{2 \pi m k_B T}}.
\end{equation}
Eq. (\ref{EquationSectionIIb}) establishes indeed a criterion that determines whether the nature of a system of particles can be considered as classical or quantum. Basically, the wavelength $\lambda_T$ serves as a length scale over which quantum effects appear. For high temperatures $\lambda_T \to 0$ and then the particles can be visualized as classical point-like particles with a definite momentum and position. However, as temperature is lowered, $\lambda_T$ starts to increase is a smooth way. There exist then a characteristic temperature $T^{*},$ such that the wavelength of particles is of the same order of magnitude as the average distance $l$ between any two particles (see Fig.\ref{FromClass2Quantum}), \emph{i.e.},
\begin{equation}\label{DegeneracyCondition}
l \simeq \lambda^*.
\end{equation}
At this temperature $T^{*},$ the system enters into the so called, \emph{degeneracy regime}. In such conditions the wave-like properties of matter drive the phenomenology of the system. Eq. (\ref{DegeneracyCondition}) is much more than a qualitative description, assembled together with Eqs. (\ref{EquationSectionIIa}) and (\ref{EquationSectionIIb}), provide the correct order of magnitude for the critical temperature of condensation $T_{c}$ in ultracold alkali gases used in current experiments of Bose-Einstein Condensation \cite{ALeggettBook}.\\

Indistinguishability of particles in the quantum regime requires the $N$-particle wave-function $\Psi(\vec{r_1},\ldots,\vec{r_N})$ of the system  satisfies certain symmetry properties. These symmetry requirements for the wave function of the $N$-particle system implies the existence of two fundamental classes of quantum systems \cite{Pauli1940}. A system for which  the total wave function is symmetric with respect to the exchange on the positions of any two particles, {\it i.e.}
\begin{equation}
 \Psi(\vec{r_1}, \vec{r_2}, \ldots, \vec{r_N} )= \Psi(\vec{r_2}, \vec{r_1}, \ldots, \vec{r_N} ),
\end{equation}
and other system where the wave function is anti-symmetric with respect to this action, i.e,
\begin{equation}\label{Antisymmetric}
\Psi(\vec{r_1}, \vec{r_2}, \ldots, \vec{r_N} )= -\Psi(\vec{r_2}, \vec{r_1}, \ldots, \vec{r_N} ).
\end{equation}
The first case corresponds to a system formed by particles called bosons while the second to a system formed by fermions. In addition, the expression (\ref{Antisymmetric}) serve as the basis for the Pauli's exclusion principle: \emph{no two identical fermions can ocupy one and the same quantum state}. Both systems, Bose and Fermi gas, exhibit completely different macroscopic properties as we shall show below.

The same symmetry considerations on the wave function has also a direct consequence on the spin of the particles involved \cite{Pauli1940}. It can be shown that for a quantum system with a symmetric wave function $\Psi$ (bosons), in the sense described above, the particle's spin $s$ can only have integer values, \emph{i.e.}, $s=0,1,2,\ldots$. For system with an antisymmetric wave function (fermions), particles can only have a spin with positive semi-integer values, that is, $s=1/2, 3/2,\ldots$. Such difference in spin values shall manifest in larger differences in their macroscopic dynamics.

\subsection{The Bose-Einstein Distribution}

In this Section we shall review the thermodynamic consequences of this symmetry condition for the chemical potential $\mu$ in an ideal Bose gas.
If we denote by $n_k$ to the number of bosons that populate a particular energy level $k$, then the symmetry principle presented above implies $n_k=0,1,2, \ldots\infty$. A direct implication of this fact is that at zero temperature, a system of Bose particles will have a macroscopic occupation of the lowest single-particle energy level. This was first recognized and discussed by Bose and Einstein in 1925 and gave rise to a large interest that culminated in the experimental realization of the first Bose-Einstein Condensate in 1995 by Ketterle \emph{et al} \cite{Ketterle} and Weiman and Cornell \cite{Cornell} with atoms of $^{26}$Na and $^{87}$Rb, respectively. The phase known as Bose-Einstein Condensate (BEC) corresponds to the state where the number of particles in the lowest-energy level $n_0$ is of the order of the total number $N$ of particles.\\

We start our discussion by noticing that two distinct situations must be considered: a case in which the number of bosons $N$ is conserved at all temperatures and the case in which is not. A well established principle in physics tells us that the number of the fermions (baryons and leptons) involved in any physical process must be conserved \cite{HalzenBook}. Thus it is expected that composite atomic bosons, such as the trapped atoms used in the BEC experiments also satisfies this principle. There are, however, many physical situations where the number of bosons in the system is intrinsically not conserved. Collective phenomena that emerge from interacting ordinary matter give rise to bosonic pseudo-particles (``quasi-particles'' or simply ``excitations''), that are created from the system's ground state by simply raising the system's temperature for instance. At equilibrium, quasi-particles are created and annihilated with a very short life-time due to thermal fluctuations. This is the main reason why they are not conserved. Interestingly, it is possible that systems composed of quasi-particles can exhibit Bose-Einstein Condensation. Indeed, experimental evidence of BEC of spin-excitations (magnon gas) \cite{MagnonsBEC} and of particle-hole coupled to photons in an optical cavity (exciton-polaritons) \cite{Rolhfing,SnokePT2010,DengRMP2010,KasprzakNature06} has been reported. The possibility of BEC in this systems relies on the fact that the quasi-particle life-time is much larger than the relaxation-time, thus the system can be described by statistical mechanics where particle number is approximately conserved.

A important system composed by bosonic particles that are not conserved is electromagnetic radiation in a cavity at thermodynamic equilibrium (black body radiation). From a quantum perspective, electromagnetic radiation can be considered as an ideal gas composed by photons that obey the rules of symmetry of the wave function $\Psi$. As in the case of quasi-particles, the number $N$ of photons is not conserved since they are continuously being absorbed and emitted by the cavity's walls. There is, however, a very recent work under progress where an experimental setup has been achieved in order to produce an electromagnetic radiation system with conserving number of particles \cite{KlaersNaturePhys2010}. In the next paragraphs we shall address the behavior of the chemical potential for a system of bosons where the number of particles is not conserved, as it occurs in the case of electromagnetic radiation confined in a cavity. In order to explore the capabilities and strengths of the ESM approach we shall partially follow the procedure suggested by Reif \cite{ReifBook}. We will find this very instructive when we extend this method to deal with quantum system with fractional statistics.

\subsubsection{The photon gas}
Let us consider a quantum ideal system composed of $N$ bosons that can be distributed along a set of $ \{ \epsilon_k \}$ energy levels. Due to the fact that we are dealing with bosons, each energy level $\epsilon_k$ might be populated with $n_k$ particles with $n_k=0,1,2, \ldots \infty$. The average number of particles $\langle n_k \rangle$ that occupies the single-particle energy level $\epsilon_k$ as a function of temperature is given by the following general prescription
\begin{equation}\label{Equationaverage}
\langle n_k \rangle= \frac{\sum_{\{n_{k}\}}n_k e^{-\beta \sum_{j}n_j \epsilon_j}}{\sum_{\{n_{k}\}} e^{-\beta \sum_j n_j \epsilon_j}},
\end{equation}
where two types of sums are defined at different levels.  For a particular macrostate $M$, there is a distribution of $n_1$ particles in the energy level $\epsilon_1$, $n_2$ particles in the energy level $\epsilon_2$ and so on. The total energy $E$ of this particular macrostate is given by
\begin{equation}\label{Equationaverage2a}
E=\sum_j n_j \epsilon_j
\end{equation}
where $j=1,2,\ldots, \infty$ if there is any possibility that the system may be excited to higher energy levels as desired. In practice the sum defined in Eq (\ref{Equationaverage2a}) ends at some finite point beyond of which there is no possibility that any particle can occupy higher energy levels. However this sum is defined only for a particular macrostate. To take into account all different possible macrostates $M$ consistent with the same total energy $E$ one must to define a second sum on a higher layer. The outer sum $\sum_{\{n_{k}\}}$ takes into account this fact and it must be performed over all possible macrostates $M$ availabe to the system. Thus in spite its apparent simplicity Eq. (\ref{Equationaverage}) is in fact a sum over the possible distributions of particles in the all possible energy levels. In order words, Eq (\ref{Equationaverage}) comprises a sum over the different sets $\{n_1, n_2, n_3, \ldots \}$ compatible with the restriction that for any of these distribution sets it must occur that the total energy $E$ of the system must be given by Eq (\ref{Equationaverage2a}).\\

The key element to perform the sum defined (\ref{Equationaverage}) is to define the partition function $Z$ of the system as
\begin{equation}
Z(V,T)=\sum_{\{n_{k}\}} e^{-\beta \sum_j n_j \epsilon_j}
\end{equation}
and then split the sum in $Z$ into the state $k$ and the remaining ones. This can be written as
\begin{eqnarray}
Z(V,T)&=&\sum_{\{n_{k}\}} e^{-\beta \left( n_1 \epsilon_1+ n_2 \epsilon_2+ \ldots+ n_k \epsilon_k+\ldots \ldots \right)}\nonumber\\
&=& \sum_{n_k} e^{-\beta n_k \epsilon_k} \sum_{n_q \neq n_k} e^{-\beta \left( n_1 \epsilon_1+n_2 \epsilon_2 + \ldots\right)}
\end{eqnarray}
where the last  sum explicitly excludes the term $n_k \epsilon_k$ which has been separated and brought up to the front of the sum. Then the quantity $\langle n_k \rangle$ can be written as
\begin{equation}\label{EquationAverage2}
\langle n_k \rangle =\frac{\sum_{n_k} n_k e^{-\beta n_k \epsilon_k} \sum_{n_q \neq n_k}e^{-\beta \left( n_1 \epsilon_1+n_2 \epsilon_2 + \ldots\right)}}{\sum_{n_k} e^{-\beta n_k \epsilon_k} \sum_{n_q \neq n_k} e^{-\beta \left( n_1 \epsilon_1+n_2 \epsilon_2 + \ldots\right)}}
\end{equation}
where the numerator has only one sum that depends on $n_k$. Up to this point these results are completely general. They do not depend on the particular values that the numbers $n_j$ can assume.\\

For bosons the numbers $n_j$ can assume any of the values $0,1,2, \ldots N$ since there is no restriction on the energy level occupancy. In the case of electromagnetic radiation and other quasi-particle systems there is no constraint on the total number of particles $N$ and thus the sums over the states $n_q \neq n_k$ in Eq (\ref{EquationAverage2}) are identical and  can be canceled out. This yields to a simpler expression for the average occupancy, i.e,
\begin{equation}
\langle n_k \rangle =\frac{\sum_{n_k}n_{k} e^{-\beta n_k \epsilon_k}}{\sum_{n_k}e^{-\beta n_k \epsilon_k}}.
\end{equation}
In order to carry out this calculation let us define the quantity $z(V,T)$ as
\begin{equation}
z(V,T)= \sum_{n_k} e^{-\beta n_k \epsilon_k}
\end{equation}
It is straightforward to see that in terms of $z(V,T)$ the average number $\langle n_k \rangle $ of occupancy can be written as
\begin{equation}
\langle n_k \rangle=-\frac{1}{\beta} \frac{\partial \ln{z(V,T)}}{\partial \epsilon_k} 
\end{equation}
Since the numbers $n_k$ can adopt any possible value between zero and infinity, the sum expressed in $z(V,T)$ is indeed a geometric series which can be readily calculated as
\begin{equation}
z(V,T)=\sum_{n_k=0}^{\infty} e^{-\beta n_k \epsilon_k}=\frac{1}{1-e^{-\beta \epsilon_k}},
\end{equation}
and thus
\begin{equation}\label{PlackDistribution}
\langle n_k \rangle = \frac{1}{e^{\beta \epsilon_k}-1}.
\end{equation}
represents the average number of particles occupying the energy level $\epsilon_k$ in the case of non conserving total number $N$ of particles. We can apply this result to the photon gas. In such case the energy of a photon with wave vector ${\bf k}$ in a given polarization is determined by $\epsilon_{k}=\hbar\omega_{\bf k}=\hbar c\vert{\bf k}\vert$. Substitution of this in in Eq. (\ref{PlackDistribution}) enables us to recover the well known \emph{Planck distribution} which gives the average distribution of photons in the $\omega_{\bf k}$ mode. It is of paramount relevance to realize that in this case $\mu$ does not appear explicitly during the calculation. We might say that chemical potential in this case is zero. However, a stronger assertion can be made. Within canonical ensembles in ESM there is no need of chemical potential if the number of particles is not conserved. \\

\subsubsection{The ideal Bose gas}

Let us now implement the above procedure for a system of bosons where the total number $N$ particles is conserved. In this category falls a vast set of system made of actual massive boson particles. The only requisite is that the total number of particles $N$ must be kept fixed.  In order to calculate the average number $\langle n_k \rangle$ in this case we must return to general expression for $\langle n_k \rangle$ given in Eq (\ref{EquationAverage2}). For the case where the number of particles is allowed to fluctuate we noticed that the sums $\sum_{n_q \neq n_k}e^{-\beta \left( n_1 \epsilon_1+n_2 \epsilon_2 + \ldots\right)}$ in the numerator and denominator were indeed the same and thus they can be canceled out. For systems where there is a restriction on the value of $N$ this is not longer the case. To understand this we just must realize that the restriction
\begin{equation}\label{EquationMar8}
 \sum_k \langle n_k \rangle = N
\end{equation}
constrains the sums given in Eq (\ref{EquationAverage2}), to be performed over the \emph{remaining} particles once the energy level $k$ has been occupied. This is, if from $N$ particles one is occupying the energy level $k$, then the remaining $N-1$ particles have to be distributed necessarily over the energy levels $q$ with $q \neq k$. This simple and powerful idea is the core of the calculation presented by Reif \cite{ReifBook}. We shall use and extended this idea to calculate averaged occupancies $\langle n_k \rangle$ beyond the BE and FD statistics.\\

For a system composed of bosons there is no restriction on the total number of particles that can occupy a single-particle energy level $\epsilon_k$. Consider for instance, the energy level $\epsilon_q$ being occupied by one particle. Then, Eq. (\ref{EquationMar8}) implies than the remaining $N-1$ particles must distribute themselves into the energy levels $k$ with $k \neq q$. In terms of Eq. (\ref{EquationAverage2}) this also implies that the sum
\begin{equation}\label{EquationSep18b}
\sum_{n_q \neq n_k}e^{-\beta \left( n_1 \epsilon_1+n_2 \epsilon_2 + \ldots\right)}
\end{equation}
must be carried out not over the all the possible values of $n_1$, $n_2$, etc but only over those that satisfy the fact that the total amount of particles available is now $N-1$ for this case and not $N$ as it was at the beginning. Let us denote this new sum as
\begin{equation}\label{EquationMar9a}
Z^{\prime}(N-1)=\sum_{n_q \neq n_k}e^{-\beta \left( n_1 \epsilon_1+n_2 \epsilon_2 + \ldots\right)},
\end{equation}
where the prime $\prime$ denotes the fact that the sum must be performed over all energy levels different form $q$. The quantity $N-1$ in parentheses indicates too that this sum is carried out over $N-1$ particles.\\

With these elements we are in position to calculate $\langle n_k \rangle$ for a system composed of $N$ bosons with $N$ fixed. For each value adopted by $n_k$, the sum in Eq(\ref{EquationMar9a}) has to be carried out over the remaining particles. This yields to
\begin{equation}\label{EquationMars9c}
\langle n_k \rangle=\frac{\displaystyle\sum_{j=0}^{N}j\,e^{-j\beta\epsilon_{k}} Z^{\prime}(N-j)}{\displaystyle\sum_{j=0}^{N}e^{-j\beta\epsilon_{k}} Z^{\prime}(N-j)}.
\end{equation}
The evaluation of Eq (\ref{EquationMars9c}) requires to compute $Z^{\prime}(j)$ from $j=1$ to $N,$ which makes the calculation rather cumbersome, see Ref. \cite{BorrmannJChemPhys93} and \cite{ReifBook} for details.  We shall detour this difficulty by taking an alternative approach to that followed by Reif. Let us $Z^{s}(N)$ from the numerator and denominator of Eq (\ref{EquationMars9c}) to obtain
\begin{equation}\label{nk}
\langle n_k \rangle=\frac{\displaystyle\sum_{j=0}^{N}j\,e^{-j\beta\epsilon_{k}} \dfrac{Z^{\prime}(N-j)}{Z^{\prime}(N)}}{\displaystyle\sum_{j=0}^{N}e^{-j\beta\epsilon_{k}} \dfrac{Z^{\prime}(N-j)}{Z^{\prime}(N)}}.
\end{equation}
Note that the ratio $Z^{\prime}(N-j)/Z^{\prime}(N)$ can be written as the product of the ratios of partition functions that differ only in one particle, \emph{i.e.},
\begin{multline}\label{ProductRatiosZ}
\frac{Z^{\prime}(N-j)}{Z^{\prime}(N)}=\frac{Z^{\prime}(N-1)}{Z^{\prime}(N)}\cdot \frac{Z^{\prime}(N-2)}{Z^{\prime}(N-1)}
\cdot\frac{Z^{\prime}(N-j)}{Z^{\prime}(N-j+1)}.
\end{multline}
We use now that the finite change of the Helmholtz free energy $\Delta F,$ when exactly just one particle is added to an $N$-particle system corresponding to a chemical potential
\begin{equation}
\mu_{N}=k_{B}T\ln\frac{Z^{\prime}(N)}{Z^{\prime}(N+1)}.
\end{equation}
With these facts,  expression (\ref{ProductRatiosZ}) can be written as
\begin{equation}\label{ProductRatiosZ2}
\frac{Z^{\prime}(N-j)}{Z^{\prime}(N)}=e^{\beta\mu_{N-1}} e^{\beta\mu_{N-2}}\cdots e^{\beta_{N-j}}.
\end{equation}
In the thermodynamic limit $N\rightarrow\infty$ we can write $e^{\beta\mu_{N-1}}=\ldots= e^{\beta_{N-j}}\approx e^{\beta\mu},$ and therefore
\begin{equation}\label{nk1}
\langle n_k \rangle=\frac{\displaystyle\sum_{j=0}^{N}j\,e^{-j\beta(\epsilon_{k}-\mu)}}{\displaystyle\sum_{j=0}^{N}e^{-j\beta(\epsilon_{k}-\mu)}}.
\end{equation}
can be readily evaluated to give the average number of particles as a function of temperature $T$ and the chemical potential $\mu$. The wanted relation reads
\begin{equation}\label{EquationSep18d}
\langle n_k \rangle = \frac{1}{e^{\beta (\epsilon_k - \mu)}-1},
\end{equation}
which is the well-known  Bose-Einstein (BE) distribution for an ideal gas of integer spin particles. In particular, the case considered here correspond to zero-spin particles. Note that taking the limit $N\rightarrow\infty$ is a crucial step to obtain properly the Bose-Einstein distribution, since our starting point is the canonical partition function of exactly $N$ particles.\\

It is worth to notice that the chemical potential $\mu$  in this context arises as a consequence of a physical restriction: the constancy of the total number of particles. From that consideration it is not obvious or straightforward to see whether the chemical potential is a positive or negative quantity. In order  $\langle n_{k}\rangle$ be a non-negative quantity  it is required that $\epsilon_{j}-\mu\le0$ for all $j$. This implies that  $\mu\le\epsilon_{0},$ where $\epsilon_{0}$ is the single-particle ground-state energy. Since in general, $\epsilon_0 \to 0$ in the thermodynamic limit, $\mu\le0$ for all temperatures. Note that the restriction imposed by Eq (\ref{EquationMar8}), gives an implicit definition of $\mu$ in terms of the particle density $n= N/V$ and temperature $T$. This is,
\begin{equation}\label{EquationSum}
N= \sum_{k=0}^{\infty} \langle n_k \rangle =  \sum_{k=0}^{\infty} \frac{1}{e^{\beta \left( \epsilon_k- \mu\right)} - 1}
\end{equation}
where the sum is strictly over an infinite number of energy levels since any particle can be occupy in principle any energy level accessible to the system.\\

To obtain the equation of state $\mu=\mu(n,T)$ implicitly defined in Eq (\ref{EquationSum}) it is convenient to transform the sum into a integral using the density of states $\rho(\epsilon)$. Given the relationship between the wave vector $k$ of a free particle contained inside a box of volume $L^3$ (under periodic boundary conditions) given by quantum mechanics $k_x=2 \pi n_x /L$ with $n_x=0, \pm 1, \pm 2, \ldots$, the sum $\sum_{k}$ over the wave vectors can be written in terms of an integral over the numbers $n_x, n_y$ and $n_z$. A change of variable enable us to write that
\begin{equation}
\sum_{k} \to \left(\frac{L}{2 \pi}\right)^3 \int dk= \frac{V}{\left( 2 \pi \right)^3} \int dk
\end{equation}
which can be represented in terms of the density of states $\rho(\epsilon)$ as
\begin{equation}
\sum_{k} \to V \int \rho(\epsilon) d \epsilon
\end{equation}
where $\rho(\epsilon)$ is a function that depends on both the system itself and its dimensionality \cite{ALeggettBook}. In three dimensions,
\begin{equation}\label{DOS}
\rho(\epsilon)= \frac{m^{2/3}}{\sqrt{2} \pi^{2} \hbar^{3}} \epsilon^{1/2}
\end{equation}
with $m$ the mass of the particles. \\

\begin{figure}[t]
\centering
\includegraphics[width=0.75\textwidth]{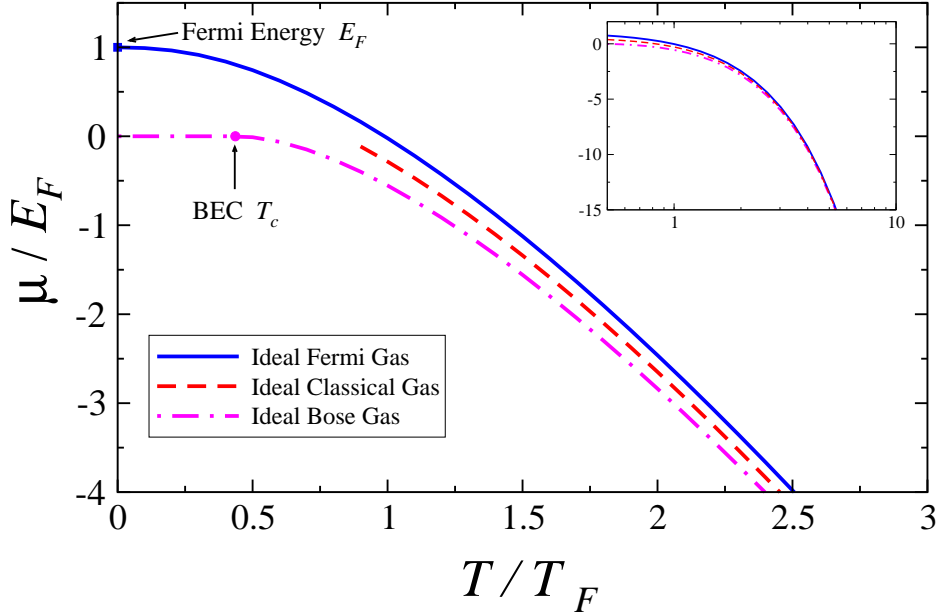}
\caption{Chemical potential in units of $E_{F}$ (the Fermi energy of a spinless ideal Fermi gas) as function of temperature in units of $T_{F}=E_{F}/k_{B}$, $k_{B}$ being the Boltzmann's constant, for: $i)$ the ideal Fermi gas (continuous-blue line), $ii)$ the ideal classical gas (red-dashed line), and $iii)$ the ideal Bose gas (magenta-dash-dotted line). The BEC critical temperature $Tc/T_{F}=[4/(3\zeta(3/2)2\sqrt{2})]^{2/3}\simeq0.436$ is marked with a dot while the Fermi Energy with a square. The inset shows how the chemical potential of the ideal quantum gas approaches the classical one at large temperatures.}
\label{BFBvsT}
\end{figure}

For free bosons in three dimensions, Equation (\ref{EquationSum}) can thus be written as an integral over the energy levels $\epsilon$ as
\begin{equation}\label{Equation23Nova}
N= \frac{V m^{3/2}}{\sqrt{2} \pi^{2} \hbar^{3}} \int_{0}^{\infty} \frac{\epsilon^{1/2} d \epsilon }{e^{\beta \left(  \epsilon- \mu \right)} - 1}
\end{equation}
where $\hbar= h/ 2 \pi$ and $V$ the volume of the system. As usual $\beta=(k_B T)^{-1}$ and $\epsilon$ is the energy of the system. The integral in Eq. (\ref{Equation23Nova}) can be expressed in terms of more familiar functions using a change of variable. For the sake of clarity we present this calculation in some detail. A new variable $x$ can be defined as $x=\beta \epsilon$. Then,
\begin{equation}\label{Equation23Novb}
\int_{0}^{\infty} \frac{\epsilon^{1/2} d \epsilon}{ e^{\beta \left( \epsilon-\mu \right)}-1}
= \frac{1}{\beta^{3/2}} \int_{0}^{\infty} \frac{x^{1/2} dx}{\zeta e^{x}-1}
\end{equation}
where $\zeta$ is defined as $\zeta \equiv e^{-\beta \mu}$. The last term in Eq (\ref{Equation23Novb}) can readily identified with a particular type of special function. This is the Poly-logarithm function $Li_{s} (\zeta)$ defined as
\begin{equation}\label{Equation23Novc}
Li_{s}(\zeta)= \frac{1}{\Gamma(s)} \int_{0}^{\infty} \frac{t^{s-1} dt}{\zeta^{-1} e^{t}-1}.
\end{equation}
For $s=3/2$ and $\zeta \to \zeta^{-1}$ we have
\begin{equation}\label{Equation23Novd}
Li_{3/2}\left(\frac{1}{\zeta}\right)= \frac{1}{\Gamma(3/2)} \int_{0}^{\infty} \frac{t^{1/2} dt}{\zeta e^{t}-1}
\end{equation}
which enable us to rewrite the last integral in Eq (\ref{Equation23Novb}) as
\begin{equation}\label{Equation23Nove}
\frac{1}{\beta^{3/2}} \int_{0}^{\infty} \frac{x^{1/2} dx}{\zeta e^{x}-1}= \frac{1}{\beta^{3/2}} Li_{3/2} \left( e^{\beta \mu}\right).
\end{equation}
With this result the chemical potential $\mu$ can be written as an implicit function of temperature $T$ and particle density $n= N/V$ as
\begin{equation}\label{Equation23Novf}
n= \frac{m^{3/2} \pi^{1/2} \left( k_B T \right)^{3/2}}{2 \sqrt{2} \pi^{2} \hbar^{3}} Li_{3/2} \left( e^{\beta \mu}\right).
\end{equation}
Equation (\ref{Equation23Novf}) corresponds to the equation of state $\mu=\mu(n,T)$ for an ideal Bose gas. This is completely equivalent to the standard equation of state for density $n$ in terms of volume $V$ and pressure $p$ as has been shown in standard thermodynamics textbooks. Both $\mu=\mu(n, T)$ and $n=n(T,p)$ contain the same information and thus can be used indistinctly to obtain thermodynamic information of the system. In Fig. \ref{BFBvsT}, the monotonic dependence on temperature of the chemical potential is shown for a fixed value of the density.

As discussed in many textbooks\cite{HuangBook,pathria} a phase transitions occurs at a critical temperature $T_{c}$ when $\mu=0.$ From Eq. (\ref{Equation23Novf}) that temperature is given by
\begin{equation}\label{Tc-BEC}
T_{c}=\frac{2\pi}{\zeta(3/2)}\frac{\hbar^{2}}{k_{B}m}n^{2/3}
\end{equation}
marked with a dot in Fig. \ref{BFBvsT}, its value in units of $T_{F}$ is given by $Tc/T_{F}=[4/(3\zeta(3/2)2\sqrt{2})]^{2/3}\simeq0.436.$ In expression (\ref{Tc-BEC}) the quantity $\zeta(3/2)=\hbox{Li}_{3/2}(1)$ is the function zeta of Riemann. In Fig. \ref{muvsrhoFermi}, isothermal curves (light-color) of $\mu(n,T)$ are shown. The critical density $n_{c}$ at which BEC occurs is determined by $\mu(n_{c},T)=0,$ and is given by $n_{c}=(mk_{B}T/2\pi\hbar^{2})^{2/2}\zeta(3/2).$\\

The peculiarity that the single-particle ground state vanishes in the thermodynamic limit can be used to discuss a thermodynamic similarity between the photon gas and the uniform ideal Bose gas. Indeed, for temperatures smaller or equal to $T_{c}$ the particles in the ground state $N_{0}$ (the condensate) do not contribute to the thermodynamics of the gas. Therefore we can disregard the condensate even when the total $N= N_{0}+N_{e}$ is fixed, where $N_{e}$ denotes the number of particles occupying the single-particle excited states. Of course, the situation described only makes sense for the ideal case and is presented here just for academic purposes. Thus, the thermodynamic properties of the gas are dictated by the behavior of $N_{e}(T)$, which grows with temperature as occurs with the photon gas in a thermal cavity. The condensate plays the role of a particle source just as the walls of the cavity emits and absorbs photons from the cavity. The quantitative difference between both systems are the result form their different dispersion relations, $\epsilon_{\bf k}=\frac{\hbar^{2}k^{2}}{2m}$ for the uniform Bose gas and $\epsilon_{\bf k}=\hbar ck$ for the photons. To exemplify this, consider the temperature dependence of the specific heat at constant volume $C_{V},$ which for $T\le T_{c}$ grows monotonically as $T^{3/2}$ for the uniform Bose gas and as $T^{3}$ for photons. This analogy would lead one to conclude that a photon gas behaves as a Bose gas with a infinite critical temperature. It is possible however to reduce the critical temperature of the photon gas to finite temperatures. One just have to manage to make the number of photons to be conserved at some critical temperature and this seems to have been recently realized experimentally by Klaers {\it et al.} by using an ingeniously experimental setup \cite{KlaersNaturePhys2010}.

\subsection{Fermi-Dirac Distribution}

Regarding fermionic systems, the electron gas has been a paramount system in solid state physics since the crucial observation of the fermionic character of the electron. Indeed, the electron gas has played a fundamental role in the first stages of the theory of metals \cite{Mott} and on the understanding of the stability of matter \cite{LiebRMP76}. After some years of the experimental realization of condensation in a degenerate Bose gas \cite{Ketterle,Cornell}, researchers started to turn their sight to the Fermi gas. The first experimental realization of a degenerate Fermi gas was carried out by de Marco and Jin \cite{DemarcoSc} exhibiting the consequences of Pauli's exclusion principle. This work has triggered a renewed interest on Fermi systems, not only  to the understanding of phenomena that emerge from strongly interacting fermion system in condensed matter, such as superconductivity, fermionic superfluidity etc., but also, to test and probe the theoretical predictions of quantum mechanics. In particular, the trapped ideal Fermi gas has been a system under intense theoretical research in the latest years \cite{VignoloPRL00,BrackPRL00,GleisbergPRA00,VignoloPRA03,Anghel03,Mueller04,AnghelJPhysA05,SongPRA06}. Many other experiments have been developed to unveil the fermionic properties of matter \cite{TruscottScience01,GranadePRL02,HadzibabicPRL03,FukuhuraPRL07}.

For particles obeying Pauli's exclusion principle, the possible values of $n_j$ are restricted to 0 and 1. In addition, a system of fermions, like the electron gas, must satisfy that total number $N$ of particles must be a constant, \emph{i.e.}, condition (\ref{EquationMar8}) must be satisfied. As mentioned before, restriction on the total number of particles implies that if a particular energy level is occupied by one particle, then the remaining $N-1$ particles should distribute themselves into different energy levels. With this as the key idea we proceed to the calculation of the average number $\langle n_k \rangle$ in the case of a Fermi-Dirac.  We tackle this calculation in a different way from what we did in the Bose-Einstein case. For the FD we shall follow closer the procedure suggested in \cite{ReifBook}.

In a similar way as we did in the BE case let us define the sum
\begin{equation}\label{EquationMar14a}
Z^{\prime}(N)= \sum_{n_q \neq n_k} e^{n_1 \epsilon_1 + n_2 \epsilon_2+\ldots}
\end{equation}
as the sum for $N$ particles carried out over all energy levels different from  $\epsilon_k$. The average number $\langle n_k \rangle$ can thus be splitted into the $k$-contribution and the remaining terms different from $k$. This is,
\begin{equation}\label{EquationMar14b}
\langle n_k \rangle =\frac{\sum_{n_k} n_k e^{-\beta n_k \epsilon_k} \sum_{n_q \neq n_k}e^{-\beta \left( n_1 \epsilon_1+n_2 \epsilon_2 + \ldots\right)}}{\sum_{n_k} e^{-\beta n_k \epsilon_k} \sum_{n_q \neq n_k} e^{-\beta \left( n_1 \epsilon_1+n_2 \epsilon_2 + \ldots\right)}}.
\end{equation}
For a system of fermions, $n_k$ can only have two values, zero or one. In addition since the number $N$ of particles is fixed once a particular energy is populated with one particle, the remaining ones must be occupied by $N-1$ particles, Eq(\ref{EquationMar14a}) together with Eq (\ref{EquationMar14b}) can be written for a fermion system as
\begin{equation}\label{EquationMar14ba}
\langle n_k \rangle=\frac{e^{\beta \epsilon_k} Z^{\prime}(N-1)}{Z^{\prime} (N)+e^{\beta \epsilon_k} Z^{\prime}(N-1)}.
\end{equation}\\

In order to relate $Z^{\prime}(N)$ with $Z^{\prime}(N-1)$ it is useful to consider the Taylor expansion of the quantity $\log{Z^{\prime}(N-\Delta N)}$ . For $\Delta N \ll N$,
\begin{equation}\label{Equationlogarithms}
\log{Z^{\prime}(N-\Delta N)} \simeq \log{Z^{\prime}(N)}- \frac{\partial \log{Z^{\prime}}}{\partial N} \Delta N.
\end{equation}
If we define $\alpha_N$ as
\begin{equation} \label{Equationalfa}
\alpha_N \equiv \frac{\partial \log{Z^{\prime}}}{\partial N},
\end{equation}
we can write Eq (\ref{Equationlogarithms}) as
\begin{equation}
\log{Z^{\prime}(N-\Delta N)} \simeq \log{Z^{\prime}(N)}- \alpha_N \Delta N,
\end{equation}
which yields to
\begin{equation}\label{EquationMar14c}
\frac{Z^{\prime} (N- \Delta N)}{Z^{\prime}(N)} = e^{-\alpha_N \Delta N}.
\end{equation}\\
Let us remember that $Z^{\prime}(N)$ is a sum defined over all states excepting the $k$ one. One may expect then that for $N \gg 1$ variations in the logarithm may be some kind of insensitive to which particular state $s$ has been omitted. Then, it may be valid that  $\alpha_N$ does not actually depends on the state $k$ chosen and thus we can simply write $\alpha_N = \alpha$ \cite{ReifBook}. Inserting this in Eq (\ref{EquationMar14c}) and performing the sum in Eq (\ref{EquationMar14ba}) accordingly,  we obtain for a fermi system the well-known Fermi-Dirac Distribution,
\begin{equation}\label{FermiDiracDistribution}
\langle n_k \rangle = \frac{1}{e^{\beta \epsilon_k + \alpha}+1},
\end{equation}
where $\alpha_s$ is given formally by Eq (\ref{Equationalfa}). A direct interpretation for $\alpha$ can be given in terms of the chemical potential $\mu$ by recalling that
\begin{equation}
\mu= \left( \frac{\partial F}{\partial N} \right)_{T,V}.
\end{equation}
Since $F=-k_BT \log{Z}$, then  $\alpha= - \mu / k_B T$. The Fermi-Dirac Distribution can be written then in a more usual form as
\begin{equation}\label{FDdistribution}
\langle n_k \rangle = \frac{1}{e^{\beta\left(  \epsilon_k - \mu \right)}+1},
\end{equation}\\
where $\mu$ is up to this point an undetermined quantity that can be obtained by imposing the following condition
\begin{equation}\label{EquationMar14d}
N= \sum_s \langle n_s \rangle = \sum_k \frac{1}{e^{\beta\left(  \epsilon_k - \mu \right)}+1}.
\end{equation}
Since both $ \langle n_k \rangle$ and $N$ should be positive quantities the chemical potential $\mu$ must adjust its value in agreement with the value of the energy levels $\epsilon_k$ in such a way that $N > 0$ and $\langle n_k \rangle \geqslant 0 $ be fulfilled in any physical situation.\\

In the same spirit as we did for our calculation in the BE case, it is possible to go a step forward to calculate explicitly the equation of state $\mu=\mu(n,T)$ for the ideal Fermi gas. By using expression (\ref{DOS}) for the density of states, Eq. (\ref{EquationMar14d}) can be written as the following integral
\begin{equation}
N= \frac{V}{\Gamma(3/2)}\left(\frac{m}{2\pi\hbar^{2}}\right)^{3/2} \int_{0}^{\infty} \frac{\epsilon^{1/2} d \epsilon }{e^{\beta \left(  \epsilon- \mu \right)} + 1},
\end{equation}
and in terms of the particle density $n$ and the polylogarithm function $Li_{s}(z)$ we have
\begin{equation}
n=- \left(\frac{mk_{B}T}{2\pi\hbar^{2}}\right)^{3/2}Li_{3/2}(-e^{\beta\mu}),
\end{equation}
which gives, implicitly, the equation of state $\mu(n,T).$
In the zero temperature limit, the FD distribution (\ref{FDdistribution}) has a step-like shape $\theta(\epsilon-\mu),$ where $\theta(x)$ is the Heaviside step function that takes the value 1 if $x\ge0$ and 0 otherwise, thus, the chemical potential $\mu(n,T=0)$ coincides with the so called Fermi energy $E_{F}=k_{B}T_{F}=\hbar^{2}k_{F}^{2}/2m$ whose dependence on $n$ is
\begin{equation}\label{FermiEnergy}
E_{F}=\frac{\hbar^{2}}{2m}\left(6\pi^{2}n\right)^{2/3}.
\end{equation}
Due to the \emph{exclusion principle} only one fermion can be allocated in a single-particle energy state (with no degeneration). Thus given $N$ particles, the system's ground state is obtained by filling the first $N$ single-particle energy states. The Fermi energy corresponds exactly to the last occupied state. For finite temperatures, but still much smaller than the Fermi temperature, the Fermi-Dirac distribution is modified from its zero temperature step-shape only around $\mu_{F}\sim E_{F}$ and the chemical potential can be computed by the use of the Sommerfeld approximation (see Ref. [\onlinecite{Ashcroft}] for details) giving the well known result
\begin{equation}\label{EquationMar14e}
\mu_{F}=E_{F}\left[1-\frac{\pi^{2}}{12}\left(T/T_{F}\right)^{2}+\ldots\right].
\end{equation}
The temperature $T^{*}$ that separates the $\mu>0$ region from the $\mu<0$ one, can be computed exactly and is given by $T^{*}=[\Gamma(5/2)\zeta(3/2)(1-\sqrt{2}/2)]^{-2/3}\, T_{F}\simeq 0.989\, T_{F},$ where $T_F$ denotes the Fermi temperature.

\begin{figure}[h]
 \includegraphics[width=0.75\textwidth]{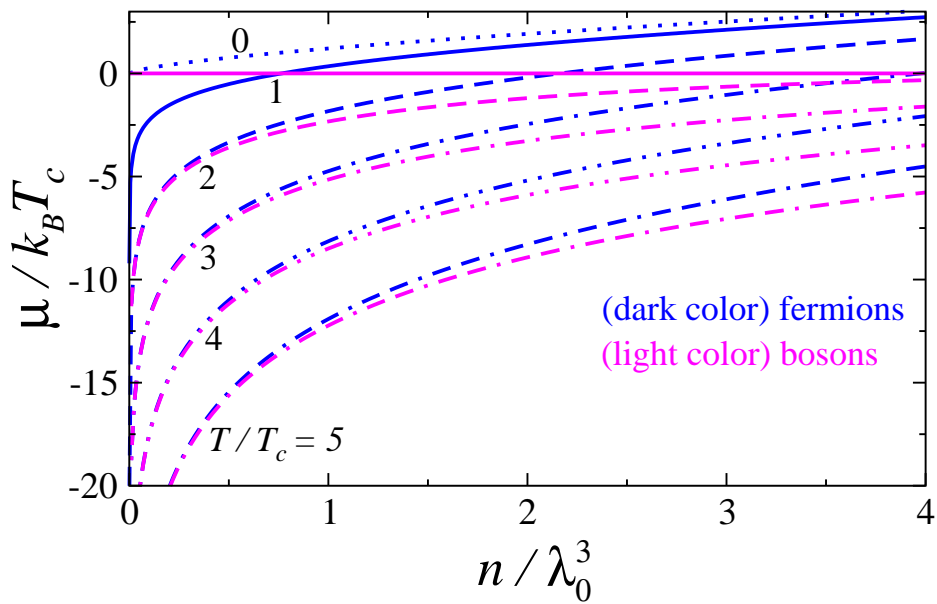}
 \caption{Chemical potential $\mu$ in units of $k_{B}T_{c}$ for the ideal Bose and Fermi gases as function of the particle density $n$ for various values of temperature. $T_{c}$ corresponds to the Bose-Einstein condensation critical temperature (\ref{Tc-BEC}) of a boson gas with the arbitrary density $n_{0}.$ $\lambda_{0}$ corresponds to the thermal wavelength evaluated at $T=T_{c}.$ Note that both cases converge to the same values of the chemical potential for small enough density, this corresponds to the classical limit. }
 \label{muvsrhoFermi}
\end{figure}

Observe that $\mu$ can be a positive quantity, even in the thermodynamic limit, in contraposition to the Bose and classical gas where it is always a negative quantity. This behavior is a direct consequence of the quantum effects at low temperatures, in this case arises from the \emph{exclusion principle}. At zero temperature, we can straightforwardly use the discrete version of Eq. (\ref{Equation22Feba}) to compute $\mu$. This is so since $\Delta S=0$ when adding exactly one particle to the system and therefore $\mu=\Delta F=\Delta U=E_{F}>0.$ In Fig. \ref{BFBvsT}, $\mu$ exhibits a monotonic decreasing dependence on temperature (blue-continuous line). Note that the transition to the classical behavior can occurs at very high temperatures, as high as the Fermi temperature which for a typical metal is of the order of 10$^{4}$ K. In Fig. \ref{muvsrhoFermi} the dependence of $\mu$ on the particle density is shown for various isotherms, for this we have chosen the scaling quantities $\mu_{0},\, n_{0}$ and $T_{0}$ of a reference system consisting of $N_{0}$ particles in the volume $V$ and Fermi energy given by (\ref{FermiEnergy}). At low densities and finite temperatures $\mu$ is negative exhibiting the classical behavior.

\section{\label{sectIV}Quantum statistics beyond Bose-Einstein and Fermi-Dirac}
As reviewed in detail in previous sections, Bose-Einstein and Fermi-Dirac statistics describe quantum systems of particles with complete different macroscopic thermodynamic effects. The essential difference between BE and FD systems is the Pauli exclusion principle which hinders the occupancy of a particular energy level to the values 0 and 1. In view of this, we address the question: Is there any intermediate case between the BE and FD statistics? Recall that both can be viewed as extreme opposites of occupancy. Whereas BE enables any number of particles from zero to $\infty$, FD blocks out any possibility beyond single-occupancy. In this Section we explore the possibility of Intermediate Quantum Statistics (IQS), i.e, statistics where any single-particle energy level can be occupied by at most $j$ particles, with $j$ an integer number between zero and $\infty$. This is the most general case with BE and FD particular cases corresponding to $j=\infty$ and $j=1$, respectively.

Let us denote with IQS$_j$, the IQS of order $j$, of a non-interacting quantum system of particles where any single-particle energy level can be occupied, at most, by $j$ particles. The calculation of the average number of particles $\langle n_k \rangle_{j} $ at the energy level $k$ corresponding to the statistics IQS$_j$ can be done in a straightforward manner by generalizing the procedure used here to calculate the BE and FD ideal statistics. As an illustrative case let us consider the calculation of $\langle n_k \rangle_{2} $ which is associated to a quantum system where the single-particle energy level $\epsilon_k$ can be occupied by zero, one or two particles. In general, as we have reviewed previously, the average number $\langle n_{k} \rangle$ can be written as
\begin{equation}\label{EquationAverage3}
\langle n_k \rangle =\frac{\sum_{n_k} n_k e^{-\beta n_k \epsilon_k} \sum_{n_q \neq n_k}e^{-\beta \left( n_1 \epsilon_1+n_2 \epsilon_2 + \ldots\right)}}{\sum_{n_k} e^{-\beta n_k \epsilon_k} \sum_{n_q \neq n_k} e^{-\beta \left( n_1 \epsilon_1+n_2 \epsilon_2 + \ldots\right)}}.
\end{equation}
where the notation is exactly the same as before. The key issue to proceed with the calculation is to realize that the two sums on the numerator and denominator in Eq (\ref{EquationAverage3}) are interrelated due to the restriction  $N=\sum_{k} \langle n_{k} \rangle$. If, for example, $n_k=1$ the sum $\sum_{n_q \neq n_k}e^{-\beta \left( n_1 \epsilon_1+n_2 \epsilon_2 + \ldots\right)}$ must be performed over the $N-1$ remaining particles since $N$ is a fixed quantity.\\

For a system obeying the IQS$_2$ statistics each energy level $\epsilon_k$ may be occupied by zero, one or two particles. Then, taking into account this $\langle n_k \rangle_{2} $ can be written explicitly as
\begin{equation}\label{Equation31Dic01}
\langle n_k \rangle_2 = \frac{e^{-\beta \epsilon_{k}} Z^{\prime}(N-1)+2 e^{-\beta \epsilon_k}
 Z^{(\prime)}(N-2)}{Z^{\prime}(N)+ e^{-\beta \epsilon_k} Z^{\prime}(N-1)+ e^{-2 \beta \epsilon_k} Z^{\prime}(N-2)}
\end{equation}
where
\begin{equation}
Z^{\prime}(N)=\sum_{n_q \neq n_k}e^{-\beta \left( n_1 \epsilon_1+n_2 \epsilon_2 + \ldots\right)}
\end{equation}
 is a sum performed over $N$ particles leaving apart the energy level $\epsilon_k$. Accordingly, $Z^{\prime}(N-1)$  represents the same sum performed over $N-1$ particles,  $Z^{\prime}(N-2)$ a sum performed over $N-2$ particles and so on. In general, $Z^{\prime}(N)$ and $Z^{\prime}(N-\Delta N)$ are related at first order by
 \begin{equation}
 Z^{\prime}(N- \Delta N)= Z^{\prime}(N) e^{-\alpha \Delta N}.
 \end{equation}
 where $\alpha$ is the fugacity and is related to the chemical potential $\mu$ by $\alpha=-\beta \mu$. Then, $\langle n_k \rangle_2$ can be written as
\begin{equation}\label{EquationIQS2}
\langle n_k \rangle_2 = \frac{e^{-\beta \left( \epsilon_k - \mu \right)}+2 e^{-2 \beta \left( \epsilon_k - \mu \right)}}{1+e^{-\beta \left(\epsilon_k-\mu \right)}+e^{-2 \beta \left(\epsilon_k - \mu \right)}}
\end{equation}
which represents the average occupancy for a quantum system with IQS$_2$ statistics. It is worth to notice that the case IQS$_1$, which represents the well-known FD statistics, is included in this expression. In such case the last terms in both the numerator and denominator are dropped out obtaining for IQS$_1$
\begin{equation}\label{Equation4Dic2010a}
\langle n_k \rangle_1 = \frac{e^{-\beta \left( \epsilon_k - \mu \right)}}{1+e^{-\beta \left(\epsilon_k-\mu \right)}}=\frac{1}{e^{\beta \left( \epsilon_k - \mu \right)}+1}
\end{equation}
which is the Fermi-Dirac Statistics. \\

The procedure outlined above can be readily generalized to calculate the average occupancy for a system with a IQS$_j$ statistics. In such case, the single-particle energy levels can be occupied by zero, one, two up to $j$ particles simultaneously. This is the most general case of a intermediate statistics between the Fermi-Dirac and Bose-Einstein cases. Please note that whereas the FD statistics corresponds to IQS$_1$, the BE statistics lies on the opposite extreme where the occupancy $j$ tends to infinity. The general expression for the average occupancy $\langle n_k \rangle_j$ in the IQS$_j$ case reads as
\begin{equation}\label{EquationEne52010a}
\langle n_k \rangle_j = \frac{\sum_{r=0}^{j} r e^{-r \beta \epsilon_{k}} Z^{\prime}(N-r)}{\sum_{r=0}^{j} e^{-r \beta \epsilon_{k}} Z^{\prime}(N-r)}
\end{equation}
which can be calculated explicitly as
\begin{equation}\label{equationEne52010b}
\langle n_k \rangle_j = \frac{e^{\left(1+j \right) \left(\alpha + \beta \epsilon_k\right)}+j-\left( 1+j \right)
e^{\left( \alpha + \beta \epsilon_k\right)}}{ \left[ e^{\left( \alpha + \beta \epsilon_k\right)} -1\right] \left[ e^{\left(1+j \right) \left(\alpha + \beta \epsilon_k\right)}-1 \right]}
\end{equation}
where $j$ can run from zero to infinity.  In order to check out that this expression is correct let us calculate some particular cases. For $j=0$ we obtain the trivial limit case with no statistics at all. If no particles are allowed to occupy any energy level then there is no average occupancy. For $j=1$ we recover the FD statistics since the expression for IQS$_1$ obtained directly from the substitution of $j=1$ in Eq (\ref{equationEne52010b})
\begin{equation}
\langle n_k \rangle_1 = \frac{e^{2 \left(\alpha + \beta \epsilon_k\right)}+1-2
e^{\left( \alpha + \beta \epsilon_k\right)}}{ \left[ e^{\left( \alpha + \beta \epsilon_k\right)} -1\right] \left[ e^{2 \left(\alpha + \beta \epsilon_k\right)}-1 \right]}
\end{equation}
is completely equivalent to Eq (\ref{Equation4Dic2010a}). The BE statistics can be also reproduced from Eq (\ref{equationEne52010b}) if we consider, as we did previously in the standard derivation of the Bose-Einstein statistics that the occupancy can run from zero to infinity. Then, the sums in Eq (\ref{EquationEne52010a}) must be carried out from zero to infinity. When that consideration is taken properly, $\langle n_k \rangle_j$ in the limit when $j \to \infty$ reproduces the BE case since
\begin{equation}\label{EquationEne52010c}
\langle n_k \rangle_{BE} = \frac{\sum_{r=0}^{\infty} r e^{-r \beta \epsilon_{k}} Z^{(s)}(N-r)}{\sum_{r=0}^{\infty} e^{-r \beta \epsilon_{k}} Z^{(s)}(N-r)}=\frac{1}{e^{\alpha+ \beta \epsilon_k}-1}
\end{equation}
where $\alpha$ as usual is the fugacity.\\

The procedure outlined and described here to calculate the average occupancy $\langle n_k \rangle_j$ in a quantum system obeying a IQS of order $j$ is based on a procedure suggested by Reif \cite{ReifBook} for the calculation of BE, FD and Planck distributions exclusively. The generalization provided here shows that the chemical potential $\mu$ and its associated quantity the fugacity $\alpha$ are physical quantities related not only to the BE and FD statistics but to all types of statistics that preserve the total number of particles. The methodology proposed here can be straightforwardly explored with undergraduate and graduate students in order to clarify how the concept of chemical potential arises and what is its role in the development of the standard FD and BE statistics. As an interesting issue to explore in this direction, it is worth to underline that once the restriction of the preservation of the number $N$ of particles is imposed this automatically restricts the summations implied in $Z^{\prime}(N), Z^{\prime}(N-1), \ldots$. All these sums are related and the connection factor is the fugacity $\alpha$ of the system. If these facts are not properly taken into account, all the sums defined by $Z^{\prime}(N), Z^{\prime}(N-1), \ldots$ may be wrongly taken as the same. This misconception will bring the cancelation of the connection factor implied.\\

To finalize this section we would like to make some comments on different approaches that have been proposed to deal with quasi-particles that are neither bosons or fermions. One of them is the concept of particles with fractional statistics also known as ``anyons'' introduced by Leinaas and Myrheim\cite{Leinaas77} and Wilczek \cite{WilczekPRL82} in two dimensional systems and that has found application in the theory of the fractional quantum Hall effect and anyon superconductivity. A completely new concept without reference to dimensionality was developed by Haldane \cite{HaldanePRL91} based on the idea that the dimension $D$ of the Hilbert space of single ``particles'' (in general quasi-particles that result from topological excitations in condensed matter) changes as particles are added to the system according to $\Delta D=-g\Delta N$. In other words, quantum correlations between ``particles'' are introduced by making the available states to depend on which states have been already occupied. The Bose statistics is recovered by setting $g=0$ and Fermi if $g=1$ \cite{WuPRL94}.

\section{\label{sectV}Weakly interacting Quantum Gases}

Let us finalize this brief review on the role of chemical potential in classical and quantum gases by briefly addressing the case of weakly interacting quantum gases. This case turns out to be of great relevance since it is the standard theoretical model to analyze Bose-Einstein Condensation in alkali atoms under magnetic and optical traps. The interacting Fermi gas , on the other hand, lies at the foundation of the superconductivity and fermionic-superfluidity theory when the effective interaction between fermions is attractive.

\subsection{The Bose Gas}

In order to describe the dynamics of a weakly interacting Bose gas it is customary to start with the general Hamiltonian operator $\hat{H}$ given by
\begin{eqnarray}\label{Equation10Ene2011a}
\hat{H} &=& \int d\vec{r}\, \hat{\Psi}^{\dagger} \left( -\frac{\hbar^{2}}{2m}+ V_{ext}(\vec{r})
\right) \hat{\Psi}+ \nonumber \\
 && \frac{1}{2}\int \int d \vec{r}
d \vec{r^{\prime}}\, \hat{\Psi}^{\dagger}(\vec{r}) \hat{\Psi}^{\dagger}
(\vec{r^{\prime}}) U(\vec{r}-\vec{r^{\prime}}) \hat{\Psi}(\vec{r})
\hat{\Psi} (\vec{r^{\prime}})
\end{eqnarray}
where $\hat{\Psi}(\vec{r})$ and $\hat{\Psi}^{\dagger}(\vec{r})$ are the field operators of annihilation and creation of particles at position $\vec{r}$ and $U(\vec{r}-\vec{r^{\prime}}) $ is the interacting potential between two particles. In general, the experimental situations involve an external potential $V_{ext}(\vec{r})$.  In the case of bosons, the field operators satisfy a particular set of commutation rules given by
\begin{equation}\label{Equation10Ene2011b}
\left[ \hat{\Psi}(\vec{r_1}), \hat{\Psi}^{\dagger}(\vec{r_2}) \right]= \delta^{3} (\vec{r_1}- \vec{r_2})
\end{equation}
and
\begin{equation}\label{Equation10Ene2011c}
\left[ \hat{\Psi}(\vec{r_1}), \hat{\Psi}(\vec{r_2}) \right]=\left[ \hat{\Psi}^{\dagger}(\vec{r_1}), \hat{\Psi}^{\dagger}(\vec{r_2}) \right]=0.
\end{equation}
The complete solution of Eq. (\ref{Equation10Ene2011a}) for any arbitrary potential $U(\vec{r}-\vec{r^{\prime}})$ is a formidable task beyond our current capabilities, however, for some particular situations it is possible to make a step further to approximate the potential $U \left( \vec{r}- \vec{r}^{\prime}\right)$ as a contact potential represented by a Dirac delta function
\begin{equation}\label{EquationEne92011}
 U \left( \vec{r}- \vec{r}^{\prime}\right)= U_{0} \delta^{3} \left( r - r^{\prime} \right),
\end{equation}
where $U_0$ is the strength of the interaction given by $U_0=  4 \pi a_s / m$ with $a_s$ the scattering length and $m$ the mass of the particle. This has proved to be particularly accurate to describe interactions in Bose gases composed of alkali atoms like $^{23}$Na, $^{87}$Rb, $^{7}$Li at very low densities and temperatures. In such systems, the interaction occurs via a $s$-wave quantum scattering process with $a_s$ the relevant parameter that characterizes the interaction between atoms.

With these considerations it is possible to rewrite Eq (\ref{Equation10Ene2011a}) as
\begin{equation}\label{Equation10Ene2011f}
\hat{H}= \sum_{q \geqslant 0} \epsilon_{q}^{0} \hat{a}^{\dagger}_{q} \hat{a}_{q}+\frac{U_{0}}{2V} \sum_{p \geqslant 0, q \geqslant 0,r \geqslant 0} \hat{a}^{\dagger}_{p+r} \hat{a}^{\dagger}_{q-r} \hat{a}_{p} \hat{a}_{q},
\end{equation}
which is a second-quantization representation \cite{FetterBook} in the momentum space $q$ for the Hamiltonian of the weakly interacting gas. The operators $\hat{a}^{\dagger}_{q}$ and $\hat{a}_{q}$ are creation and annihilation operator \emph{in the momentum space}. The Hamiltonian in Eq. (\ref{Equation10Ene2011f}) can be split up into the zero momentum state $q=0$ and states with $q \neq 0$. Neglecting terms of the order $N^{-1}$ which vanish in the thermodynamic limit, Eq. (\ref{Equation10Ene2011f}) is written as,
\begin{equation}\label{Equation2.2}
H=\frac{N_{0}^{2} U_{0}}{2 V}+ \sum_{q \neq 0} \left( \epsilon_{q}^{0}+ 2 n_{0}U_{0}\right)\hat{a}_{q}^{\dagger} \hat{a}_{q} + \frac{U_{0}}{V} \sum_{p,q} \hat{a}_{p}^{\dagger}\hat{a}_{p} \hat{a}_{q}^{\dagger} \hat{a}_{q},
\end{equation}
where  $n_{0}= N_{0} {\Psi_{0}}^{2}$ is the density of the condensate and $\Psi_{0}$ is the corresponding wave function. This Hamiltonian can be expanded around an equilibrium occupation distribution $f_{q}$ which for bosons is the Bose-Einstein distribution function \cite{Pethick}. To first order, the Hamiltonian is
\begin{equation}\label{Equation2.25}
H=\frac{N_{0}^{2} U_{0}}{2 V}+ \sum_{q \neq 0} \left( \epsilon_{q}^{0}+ 2 n U_{0}\right)\hat{a}_{q}^{\dagger} \hat{a}_{q} - \frac{U_{0}}{V}\sum_{p, q}f_{p}f_{q},
\end{equation}
where $n= n_{0} + n_{1}$ is the total particle density of the system and $n_{1}=\sum_{q >0} N_{q} \Psi_{q} ^{2}$ is the density of uncondensed particles which is a sum over all the non zero momentum states.  The Hamiltonian in Eq (\ref{Equation2.25}) is known as the Hartree-Fock (HF) approximation for the weakly interacting Bose gas. The second term of Eq (\ref{Equation2.25}) shows the intrinsic nature of the Hartree-Fock approximation as a mean field theory. The energy $\epsilon_{q}^{0}+ 2 n U_{0}$ to add or remove  a particle to a state with non zero momentum is an average over all the pairwise interactions between particles.

The equations for the wave functions of the condensate $\Psi_{0}( \vec{r})$ and the uncondensed phase $\Psi_{k} (\vec{r})$ can be obtained from the Heisenberg Equation $ -i / \hbar \left[ H, \Psi \right]= \partial_{t} \Psi$  with $H$ given by Hartree-Fock approximation in Eq (\ref{Equation2.25}) as

\begin{equation}\label{Equation2.3}
\begin{split}
-\frac{\hbar^{2}}{2m}\nabla^{2} \Psi_{0}+2 n_{1}U_{0} \Psi_{0}+n_{0} U_{0} \Psi_{0}+ V_{ext} \Psi_{0}= \epsilon_{0} \Psi_{0} \\
-\frac{\hbar^{2}}{2m}\nabla^{2} \Psi_{q}+2 U_{0} \left[ n_{1} + n_{0}  \right] \Psi_{q}+ V_{ext} \Psi_{q}= \epsilon_{q} \Psi_{q}
\end{split}
\end{equation}
with the last equation valid for $q\neq0$.
Since $N_{0}$ and $N_{k}$ for $k >0$ are assumed to obey a Bose-Einstein statistics, Eqs. (\ref{Equation2.3}) enable us to obtain the thermodynamic framework of the interacting Bose gas in the HF approximation. These equations determine the chemical potential $\mu$ in terms of the total particle density $n$ and the temperature $T$ as
\begin{equation}\label{Equation2.4}
n \lambda_{T}^{3} = g_{3/2} \left[ \beta \left( \mu-2 n U_{0}\right) \right]
\end{equation}
for $ T >T_{c}$ and
\begin{equation}
n= n_{0}+\frac{1}{\lambda_{T}^{3}}g_{3/2} \left( - \beta n_{0} U_{0} \right)
\end{equation}
with $\mu=U_{0}(2n -n_{0})$ for $T < T_{c}$, with
\begin{equation}\label{Equation2.5}
g_{v} (\alpha) =\frac{1}{\Gamma (v)} \int_{0}^{\infty} \frac{x^{v-1}}{e^{x - \alpha}-1}dx
\end{equation}
is the Bose integral, and $\lambda_{T}$ is the thermal de Broglie wavelength (\ref{EquationSectionIIb}).
For a system of units where $a_{s}=m=\hbar=k_{B}=1$, Eqs. (\ref{Equation2.4}) can be written in  dimensionless form as,
\begin{equation}\label{Equation2.7a}
n= \left( \frac{T}{2 \pi} \right)^{3/2} g_{3/2} \left[ \frac{\mu-8 \pi n}{T}\right]
\end{equation}
for $ T >T_{c}$ and
\begin{equation}\label{Equation2.7b}
n= n_{0} + \left( \frac{T}{2 \pi}\right)^{3/2} g_{3/2} \left( - \frac{4 \pi n }{T}\right)
\end{equation}
with $ \mu= 4 \pi \left( 2 n - n_{0} \right)$ for $ T < T_{c}$. Here the strength of the interaction $U_0$ has been replaced by its dimensionless $4 \pi$ value.\\

\begin{figure}[ht]
\centering
\includegraphics[width=0.75\textwidth]{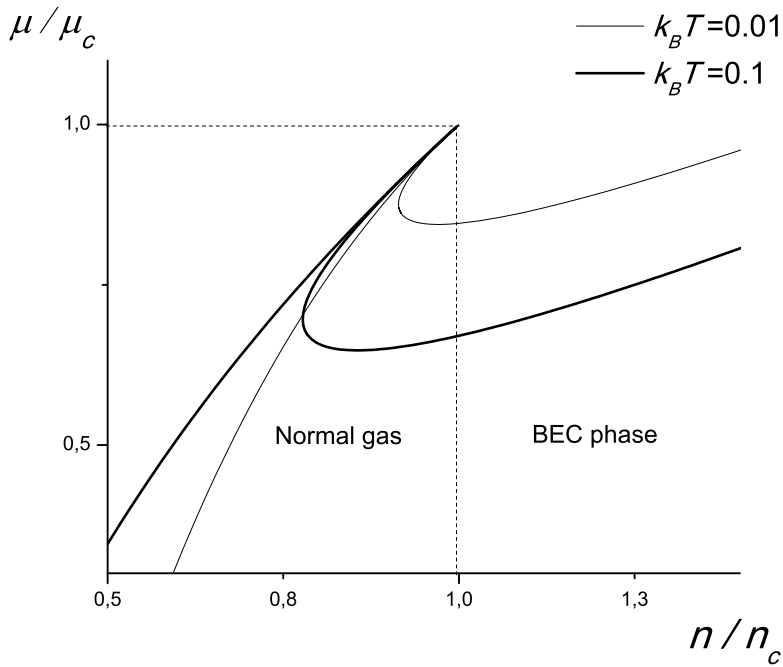}
\caption{Chemical potential $\mu$ as function of total density $n$ in the Hartree-Fock approximation for two fixed temperatures $k_B T=0.01$ and $k_B T=0.1$ in units where $m=\hbar=a_s=1$. For temperatures $T$ above the critical temperature $T_c$, i.e, densities $n$ below the critical density $n_c$ the HF approximation gives correctly the behavior of the chemical potential for which $\mu \to -\infty$. However, for $T < T_c$ or equivalently for $n > n_c$, the HF approximation yields to a chemical potential with a non single-value behavior. This is unacceptable based on fundamental principles of thermodynamics.}
\label{HFApproximation2Plots}
\end{figure}

Eqs. (\ref{Equation2.7a}) and (\ref{Equation2.7b}) contain all the relevant thermodynamic information for the weakly interacting Bose gas in the HF approximation and it is possible to solve them for $\mu$ in terms of the total particle density $n$ and temperature $T$ in order to obtain the isotherms of the equation of state $\mu=\mu(n,T)$ for a gas confined in a box of volume $V$. Recently one of us has address this issue
\cite{OlivaresRomeroJPB} for different values of the gas parameter $\gamma= {a_s}^{3} n$ obtaining isotherms for the weakly interacting Bose gas in the HF approximation (Fig. \ref{HFApproximation2Plots}). The results show that the HF approximation while a valid theory of the interacting gas near zero temperature fails to predict and adequate physical behavior near the transition. Indeed, in the vicinity of the critical density $n_c$, the HF formalism predicts a non single-valued profile for $\mu(n)$ a feature forbidden by fundamental thermodynamic principles.

\subsection{Interacting Fermi gas}

The corresponding Hamiltonian for fermions with two spin states $\sigma=\uparrow,\, \downarrow$, as naturally occurs in several condensed matter systems, may be written as
\begin{multline}\label{FermiFieldsHamiltonian}
\hat{H} = \sum_{\sigma}\int d{\bf r}\, \hat{\Psi}^{\dagger}_{\sigma}({\bf r}) \left(-\frac{\hbar^{2}\nabla^{2}}{2m}+ V_{ext}(\sigma,{\bf r})-\mu_{\sigma}
\right) \hat{\Psi}_{\sigma}({\bf r}) \\
  +\int d{\bf r}d{\bf r^{\prime}}\,  \hat{\Psi}^{\dagger}_{\uparrow}({\bf r})\hat{\Psi}^{\dagger}_{\downarrow}
({\bf r}^{\prime}) U({\bf r}-{\bf r^{\prime}}) \hat{\Psi}_{\downarrow}({\bf r})\hat{\Psi}_{\uparrow}({\bf r^{\prime}})
\end{multline}
where the field operators obey the fermionic anticommutation relations $\{\hat{\Psi}_{\sigma}({\bf r}),\hat{\Psi}^{\dagger}_{\sigma^{\prime}}({\bf r}^{\prime})\}\equiv \hat{\Psi}_{\sigma}({\bf r})\hat{\Psi}^{\dagger}_{\sigma^{\prime}}({\bf r}^{\prime})+\hat{\Psi}^{\dagger}_{\sigma^{\prime}}({\bf r}^{\prime})\hat{\Psi}_{\sigma}({\bf r})=\delta_{\sigma,\sigma^{\prime}}\delta({\bf r}-{\bf r}^{\prime}).$
The density of fermions in each spin state $n_{\uparrow,\downarrow}$ fixes the energy scale $E_{F}$ of the noninteracting fermion given by expression (\ref{FermiEnergy}) with $n=\frac{1}{2}n_{\uparrow}=\frac{1}{2}n_{\downarrow}$.

Two different aspects of the interacting Fermi gas are obtained depending on whether the interaction between fermions is attractive or repulsive. \emph{At zero temperature} corrections to the ideal case value of the chemical potential can be calculated in terms of the scattering length $a_s$ which measures the interaction strength. In the dilute limit we have $k_{F}a_s\ll1$, where the Fermi wavevector $k_{F}$ depends on the particle density as in (\ref{FermiEnergy}).\\

In the repulsive case there is no change in the intuition we have developed from the interacting classical gas. So one would expect the chemical potential to rise above the value of the noninteracting case. This is clear from the Landau theory of the Fermi liquid \cite{NozieresBook}. A calculation due to Galitskii (see Ref. [\onlinecite{FetterBook}] pp. 147) gives that for $k_{F}a$ sufficiently small, the chemical potential can be written as
\begin{equation}
\mu= \frac{\hbar^{2}k_{F}^{2}}{2m}\left[1+\frac{4}{3\pi}k_{F}a_s+\frac{4}{15\pi^{2}}(11-2\ln2)(k_{F}a_s)^{2}\right]
\end{equation}
exhibiting that repulsive interaction leads to an increase of the chemical potential as it occurs in the classical case. There is a particular interest in the case when attractive interactions between fermions of different spin-polarization are considered. This is, for example, the case in superconductors or in ultracold Fermi gases manipulated through magnetic fields, where the possibility of Cooper pairing is present. The formation of Cooper pairs, even for a extremely weak attraction, makes the Fermi sea unstable \cite{CooperPR56} leading to a ground state different from the Fermi liquid one called the Bardeen-Cooper-Schrieffer (BCS) ground-state. This microscopic mechanism developed further by Bardeen, Cooper and Schrieffer\cite{BardeenPR57} serves as the basis for the explanation of conventional superconductivity.

In the weak coupling limit, the chemical potential at zero temperature does not deviate significantly from the the noninteracting case value $E_{F}$. This picture changes if the strength of the attractive potential is increased and $\mu$ is computed in a self-consistent theory called the BCS-BEC crossover. Such a theory, introduced by Eagles\cite{EaglesPR69} and Leggett\cite{Leggett80}, extends the BCS one, where now the chemical potential changes due to the formation of tightly bound fermion pairs. The BCS-BEC crossover has been developed and widely applied, first in the context of high-Tc superconductivity and most recently in the formation of fermionic molecules in ultracold Fermi gases. We restrict our discussion to the case of zero temperature giving references for the finite temperature case.

The Hamiltonian (\ref{FermiFieldsHamiltonian}) can be written in momentum space as
\begin{multline}\label{FermionHamiltonian}
\hat{H}= \sum_{{\bf k},\sigma} (\epsilon_{\bf k}-\mu) c^{\dagger}_{{\bf k},\sigma} c_{{\bf k},\sigma}
-\sum_{{\bf k},{\bf k}^{\prime},{
\bf q}}V_{k,k^{\prime}} c^{\dagger}_{{\bf k+q/2},\uparrow} c^{\dagger}_{{\bf -k+q/2},\downarrow} c_{{\bf k+q/2},\downarrow} c_{{\bf -k+q/2},\uparrow},
\end{multline}
where $V_{k,k^{\prime}}$ is the two-body attractive interaction. In a self-consistent mean-field theory, the order parameter $\Delta_{k} = \sum_{\bf k^{\prime}}V_{k,k^{\prime}}\langle c^{\dagger}_{{\bf k},\uparrow} c^{\dagger}_{{\bf -k},\downarrow}\rangle$ called the "gap" obeys the well known gap equation
\begin{equation}\label{GapEquation}
\Delta_{k}=\sum_{\bf k^{\prime}}V_{k,k^{\prime}}\frac{\Delta_{k^{\prime}}}{2(\epsilon_{\bf k}-\mu)}(1-2n_{k^{\prime}})
\end{equation}
where
\begin{equation}\label{nkBCScrossover}
n_{k}=\frac{1}{2}\left\{1-\frac{(\epsilon_{\bf k}-\mu)}{[(\epsilon_{\bf k}-\mu)^{2}+\Delta_{k}^{2}]^{1/2}}\right\},
\end{equation}
gives the quasi-particle density with wavevector $k$. The simultaneous solution of these equations give $\mu$ and $\Delta$ as function of the strength of the interfermionic interaction.\\

The BEC-BCS crossover has implications on the behavior of $\mu$ as a function of the interaction strength \cite{ChenPhysRep2005}. In the weak coupling regime, $\mu= E_{F}$ and ordinary BCS theory applies. At sufficiently strong coupling, $\mu$ starts a monotonous decreasing behavior with increasing the attraction strength, eventually it crosses the zero value and then becomes negative in the Bose regime. With the appearance of tightly-bound pairs but $\mu$ still positive, the system has a remnant of the Fermi surface, and we say that the many-body system preserves a fermionic character. For negative $\mu,$ however, no trace of a Fermi surface is left and the system is considered rather bosonic.

\section{\label{sectVI}Conclusions and Final Remarks}
The concept of chemical potential in the context of classical and quantum gases has been revisited.
For the classical gas, we started on giving a physical argument on why one should expect $\mu<0$ and then we considered the effects of the inter-particle interactions for the particular case of the van der Waals gas.
Based on a equivalence due to Widom that relates the work necessary to bring an extra particle from infinity to a given position in the system, with the difference in the chemical potential respect to the perfect gas, we give a physical argument on what to expect when interactions are considered. Thus, in the case of a pure repulsive interaction of the hard-sphere type of radius $d$, the chemical potential is shifted above the ideal gas value by the amount $\frac{4}{3}\pi (d/l)^{3}k_{B}T.$ This result is valid only in the low density limit, {i.e.}, $d\ll l$. For this case it is intuitively clear, that in the situation just described, it must be spent more energy in bringing an extra particle to the system than when no interactions are present at all and that this amount of energy increases as the density does. Thus an infinite amount of energy will be required to add a particle to a high dense classical gas. As the reader can now expect, attractive interactions shifts the chemical potential below the value of the non-interacting case. This is the case when considering the attractive tail in the model potential (\ref{vanWpotential}).\\

For the ideal quantum gases, we present a pedagogical way to obtain the Bose-Einstein and Fermi-Dirac distributions starting from a canonical-ensemble calculation. In the case of bosons, we discussed the implications on the chemical potential when considering a system, both, with conserving and non-conserving number of particles. In the later case, no reference to $\mu$ is needed, however, $\mu$ appears naturally once the number of bosons is required to be conserved. In this case $\mu(T)$ decreases monotonically with temperature lying below, but asymptotically approaching, to the classical curve (see Fig. \ref{BFBvsT}). In contrast, for fermions it was shown that $\mu$ acquire positive values due to the statistical correlations induced by Pauli's exclusion principle. In addition, $\mu(T)$ decreases monotonically from the Fermi energy lying above, but asymptotically approaching, to the classical curve (see Fig. \ref{BFBvsT}). On the light of these observations, we can use the ideas exposed for the interacting classical gas. Indeed, if we consider the quantum gas as classical, with quantum correlations given by a \emph{statistical} interparticle potential $v_{stat,ij}$ \cite{pathria,UhlenbeckPR32}, then, due to the attractive/repulsive nature of $v_{stat,ij}$ for the Bose/Fermi gas, their respective chemical potentials vary with temperature below/above the classical one.\\

We have also briefly discussed the consequences of considering an extension of the exclusion principle when a single-particle energy level can be occupied at most for $j$ particles. Finally, we have presented a discussion on the behavior of $\mu$ for the case of weakly interacting quantum gases. For the Bose gas, $\mu(T,n)$ gives important information on the nature of the BEC phase transition. In the case of the attractively interacting Fermi gas, $\mu$ gives important information on the nature of the system as the interaction strength is varied, going from loosely bound pairs (Cooper pairs) in the weak coupling to bosonic thightly bound-pairs in the strong interaction limit.

\section{\label{AppendixOne} Appendix}
The basic idea to evaluate the configurational integral $Q_{N}$ given by Eq. (\ref{Qconfigurational}), is to compute the statistical average of $e^{-\beta v_{1,2}}e^{-\beta v_{1,3}}\cdots e^{-\beta v_{N-1,N}}$ over all possible configurations of the particles positions denoted with $Q_{N}/V^{N}=\overline{e^{-\beta v_{1,2}}e^{-\beta v_{1,3}}\cdots e^{-\beta v_{N-1,N}}}$. van Kampen's approach is based on a factorization of $D$ into terms $D_{k}$ that takes into account the correlations of $k\ge2$ particles, \emph{i.e.},
\begin{equation}\label{ConfigurationalInt2}
Q_{N}/V^{N}=\prod_{k=2}^{N}\left(d_{k}\right)^{\binom{N}{k}},
\end{equation}
where $\binom{N}{k}$ gives the number of combinations of $k$ particles taken from the total $N,$ and
\begin{equation}\label{kthFactor}
d_{k}=\frac{\overline{e^{-\beta v_{1,2}}e^{-\beta v_{1,3}}\cdots e^{-\beta v_{k-1,k}}}}{D},
\end{equation}
with $D$ is the immediate lower approximation for the same numerator. 

For $k=2,$ $d_{2}=\overline{e^{-\beta v_{1,2}}}$ since $D=1$ in this case. Thus the first factor in Eq. (\ref{ConfigurationalInt2}) is given by
\begin{equation}
\overline{e^{-\beta v_{1,2}}}^{N(N-1)/2}=\left[V^{-1}\int d{\bf r}_{1}V^{-1}\int d{\bf r}_{2}\, e^{-\beta v_{1,2}}\right]^{N(N-1)/2}.
\end{equation}
In order to take the thermodynamic limit $N,V\rightarrow\infty$ with $N/V$ constant, consider the following identity
\begin{multline}
 \left[\int \frac{d{\bf r}_{1}}{V}\int \frac{d{\bf r}_{2}}{V}\, e^{-\beta v_{1,2}}\right]^{(N-1)/2}\\
 = \left[1+\frac{1}{N}\frac{N}{V}\int d{\bf r}\, \left(e^{-\beta v({\bf r})}-1\right)\right]^{(N-1)/2},
\end{multline}
thus giving as result $d_{2}^{\binom{N}{2}}=\exp\left\{\frac{N^{2}}{2V}B_{1}\right\}$
with $B_{1}\equiv\int d{\bf r}\, \left(e^{-\beta v({\bf r})}-1\right).$ For dilute enough systems where only correlations of two particles are important this approximation should work fine.

The calculation of the general factor $\left(d_{k}\right)^{\binom{N}{k}}$ is more involved and we present only a sketch of it. By writing $e^{-\beta v_{i,j}}=1+f_{i,j}$ Eq. (\ref{kthFactor}) can be rewritten as
\begin{equation}\label{kthFactor2}
d_{k}=\frac{1+\overline{f_{1,2}}+\ldots+\overline{f_{1,2}f_{1,3}\cdots f_{k-1,k}}}{D}.
\end{equation}
van Kampen argues that the class of terms in the numerator of (\ref{kthFactor2}) that involve less than $k$ particles and those that involve $k$ particles but are reducible, are also present in $D,$  such that the numerator can be written as $(1+\sum_{\{k\}}\overline{f_{1,2}f_{1,3}\cdots}+\mathcal{O}(V^{-k}))D$, where the summation extends over all irreducible terms that involve $2,\ldots,k$ particles. A term of the form $\int\cdots\int\prod_{i<j}g_{i,j}\, d{{\bf r}_{1}}\cdots d{{\bf r}_{k}}$, with $g_{i,j}$ an arbitrary function of $\vert{\bf r}_{i}-{\bf r}_{j}\vert,$ is said to be irreducible if cannot be factorized into products of integrals of $g_{i,j}$ involving less than $k$ particle-position coordinates. For instance, it is straightforward to check that $\iiint d{\bf r}_{1}d{\bf r}_{2}d{\bf r}_{3}\, g_{1,2}g_{1,3}g_{2,3}$ is irreducible while the integral $\iiint d{\bf r}_{1}d{\bf r}_{2}d{\bf r}_{3}\, g_{1,2}g_{1,3}=V\left[\int d{\bf r}g({\bf r})\right]^{2}$ is not, here we have used the identity $\int d{\bf r}=V.$

Thus, we have
\begin{equation}\label{kthFactor2}
d_{k}=1+\frac{(k-1)!}{V^{k-1}}B_{k-1}+\mathcal{O}(V^{-k}),
\end{equation}
where $B_{k}$ is given by (\ref{B_k}) and we have recognized $\sum_{\{k\}}\overline{f_{1,2}f_{1,3}\cdots}$ with the usual irreducible cluster integral \cite{HuangBook,pathria,ReifBook} $\frac{(k-1)!}{V^{k-1}}B_{k-1}.$ In the thermodynamic limit the factor $\left(d_{k}\right)^{\binom{N}{k}}$ can then be written as
\begin{equation*}
\exp{\left\{\frac{N^{k}}{V^{k-1}}\frac{B_{k-1}}{k}\right\}}
\end{equation*}
and by combining this result with the result for $k =2$ we finally get the desired result given by expression (\ref{RealGasZ2}).

\begin{acknowledgements}
FJS akcnowledge partial support from the DGAPA grant PAPIIT-IN117010. L. Olivares-Quiroz would like to acknowledge partial support from Universidad Autonoma de la Ciudad de Mexico.
\end{acknowledgements}

\end{document}